\documentclass[12pt]{article}

\usepackage{epsf}
\usepackage{latexsym,amssymb,euscript}
\usepackage[dvips]{graphicx}
\usepackage{amsmath}

\begin{document}

\begin{center}
{\Large \textbf{Phase transitions in strongly coupled three-dimensional
$Z(N)$ lattice gauge theories at finite temperature}}
\end{center}

\vskip 0.3cm
\centerline{O.~Borisenko$^{1\dagger}$, V.~Chelnokov$^{1*}$, 
G.~Cortese$^{2\dagger\dagger}$, R.~Fiore$^{3\P}$,
M.~Gravina$^{4\ddagger}$, A.~Papa$^{3\P}$, I.~Surzhikov$^{1**}$}

\vskip 0.6cm

\centerline{${}^1$ \sl Bogolyubov Institute for Theoretical Physics,}
\centerline{\sl National Academy of Sciences of Ukraine,}
\centerline{\sl 03680 Kiev, Ukraine}

\vskip 0.2cm

\centerline{${}^2$ \sl Instituto de F\'{\i}sica Te\'orica UAM/CSIC,}
\centerline{\sl Cantoblanco, E-28049 Madrid, Spain}
\centerline{\sl and Departamento de F\'{\i}sica Te\'orica,}
\centerline{\sl Universidad de Zaragoza, E-50009 Zaragoza, Spain}

\vskip 0.2cm

\centerline{${}^3$ \sl Dipartimento di Fisica, Universit\`a della 
Calabria,}
\centerline{\sl and Istituto Nazionale di Fisica Nucleare, 
Gruppo collegato di Cosenza}
\centerline{\sl I-87036 Arcavacata di Rende, Cosenza, Italy}

\vskip 0.2cm

\centerline{${}^4$ \sl Department of Physics, University of Cyprus,
P.O. Box 20357, Nicosia, Cyprus}

\vskip 0.6cm

\begin{abstract}
We perform an analytical and numerical study of the phase transitions in 
three-dimensional $Z(N)$ lattice gauge theories at finite temperature for 
$N>4$ exploiting equivalence of these models with a generalized version 
of the two-dimensional vector Potts models in the limit of vanishing 
spatial coupling. In this limit the Polyakov loops play the role of $Z(N)$ spins.
The effective couplings of these two-dimensional spin models are calculated 
explicitly. It is argued that the effective spin models have two phase 
transitions of BKT type. This is confirmed by large-scale Monte Carlo 
simulations. Using a cluster algorithm we locate the position of the critical 
points and study the critical behavior across both phase transitions in 
details. In particular, we determine various critical indices, compute the 
helicity modulus, the average action and the specific heat. 
A scaling formula for the critical points with $N$ is proposed.
\end{abstract}

\vfill
\hrule
\vspace{0.3cm}
{\it e-mail addresses}:

$^\dagger$oleg@bitp.kiev.ua, \ \ $^*$vchelnokov@i.ua,
\ \ $^{\dagger\dagger}$cortese@unizar.es,
\ \ $^{\P}$fiore, papa \ @cs.infn.it, 

$^{\ddagger}$gravina@ucy.ac.cy,\ \ $^{**}$i\_van\_go@inbox.ru

\newpage

\section{Introduction}

The phase structure of three-dimensional ($3d$) pure $Z(N)$ lattice gauge 
theories (LGTs) has been the subject of an intensive study for more than three 
decades. It is well known by now that the zero-temperature models possess a 
single phase transition which disappears in the limit 
$N\to\infty$~\cite{bhanot}. Thus, the $U(1)$ LGT has a single confined phase 
in agreement with theoretical results~\cite{3d_u1}.  
The deconfinement phase transition at finite temperature is well understood 
and studied for $N=2,3$. These models belong to the universality class of 
$2d$ $Z(N)$ spin models and exhibit a second order phase transition in 
agreement with the Svetitsky-Yaffe conjecture~\cite{svetitsky}. Much less is 
known about the finite-temperature deconfinement transition when $N>4$. 
The Svetitsky-Yaffe conjecture is known to connect critical properties 
of $3d$ $Z(N)$ LGTs with the corresponding properties of $2d$ spin models,
if they share the same global symmetry of the action. It is widely expected, 
and in many cases proved by either analytical or numerical methods, that some 
$2d$ $Z(N>4)$ spin models (like the vector Potts model) have two phase 
transitions of infinite order, known as the Berezinskii-Kosterlitz-Thouless 
(BKT) phase transitions. Then, according to the conjecture, the phase 
transitions in some $3d$ $Z(N>4)$ gauge models at finite temperature could 
exhibit two phase transitions as well. 
Moreover, if the correlation length diverges when approaching the critical 
point, these transitions should be of the BKT type and belong to the 
universality class of the corresponding $2d$ $Z(N)$ spin models.

The BKT phase transition is known to take place in a variety of $2d$ systems: 
certain spin models, $2d$ Coulomb gas, sine-Gordon model, Solid-on-Solid 
model, etc.
The most elaborated case is the $2d$ $XY$ 
model~\cite{berezin,kosterlitz1,kosterlitz2}.
There are several indications that this type of phase transition is not a rare
phenomenon in gauge models at finite temperature - one can argue that in some
$3d$ lattice gauge models the deconfinement phase transition is of
BKT type as well. A well known example is the deconfinement phase transition 
in the $U(1)$ LGT. Indeed, certain analytical~\cite{svetitsky,parga,lat_07} 
as well as numerical results~\cite{3du1ft,3du1full} unambiguously indicate the 
BKT nature of the phase transition\footnote{It should be noted, however, that 
the numerical results of~\cite{3du1full} point to a critical index~$\eta$
larger than its $XY$ value by almost a factor of 2 for $N_t=8$. 
Therefore, the question of the universality remains open for this model.}. 

Many details of the critical behavior of $2d$ $Z(N)$ spin models are well known
-- see the review in Ref.~\cite{Wu}. The $Z(N)$ spin model in the Villain
formulation has been studied analytically in
Refs.~\cite{elitzur,savit,kogut,nienhuis,kadanoff,Cardy}. It was shown that 
the model has at least two phase transitions when $N\geq 5$. The intermediate 
phase is a massless phase with power-like decay of the correlation function. 
The critical index $\eta$ has been estimated both from the renormalization 
group (RG) approach of the Kosterlitz-Thouless type and from the 
weak coupling series for the susceptibility.
It turns out that $\eta(\beta^{(1)}_{\rm c})=1/4$ at the transition point 
from the strong coupling (high-temperature) phase to the massless phase, 
{\it i.e.} the behavior is similar to that of the $XY$ model. At the 
transition point $\beta^{(2)}_{\rm c}$ from the massless phase to the ordered 
low-temperature phase one has $\eta(\beta^{(2)}_{\rm c})=4/N^2$.
A rigorous proof that the BKT phase transition does take place, and so that the
massless phase exists, has been constructed in Ref.~\cite{rigbkt} for both
Villain and standard formulations of the vector Potts model. 
Universality properties of vector Potts models were studied via Monte Carlo 
simulations in Ref.~\cite{cluster2d} for $N=6,8,12$ and in 
Refs.~\cite{lat_10,2dzn,lat_11,2dzn7_17} for $N=5,7,17$. 
Results for the critical indices $\eta$ and $\nu$ agree well with the
analytical predictions obtained from the Villain formulation of the model.

We expect that $3d$ $Z(N)$ gauge models at finite temperature exhibit
the same critical properties as $Z(N)$ spin models in two dimensions.
In particular, gauge models with $N>4$ may possess two phase transitions
of the BKT type. On the basis of the Svetitsky-Yaffe conjecture
the critical behavior of the gauge model in this case is governed by the $2d$
$Z(N)$ spin model. In particular, one expects the following values 
of critical indices: $\nu = 1/2,\ \eta = 1/4$ at the first transition
and $\nu = 1/2,\ \eta = 4/N^2$ at the second transition.
To the best of our knowledge this scenario was not verified in the literature 
by either analytical or numerical means. The main goal of the present paper is
to fill this gap and to study the nature of deconfining phase transitions in 
$3d$ $Z(N)$ LGTs. 

The fact that the BKT transition has infinite order makes it hard to study 
its properties using analytical methods. In most of the cases studied one 
uses a renormalization group (RG) technique like in Ref.~\cite{elitzur}. 
Unfortunately, there are no direct ways to generalize transformations 
of Ref.~\cite{elitzur}, leading to RG equations, to $3d$ $Z(N)$ LGTs except 
for the limiting case $N\to\infty$. To study the phase structure of these 
models we need numerical simulations. Here, however, another problem appears
related to the very slow, logarithmic convergence to the thermodynamic limit 
in the vicinity of the BKT transition. It is thus necessary to use both 
large-scale simulations and combine them with the finite-size scaling methods. 
For a full finite-temperature gauge model this is an ambitious program, 
especially if one wants to study several values of $N$. 
We have therefore decided to utilize an approach developed by some of the
present authors in Refs.~\cite{3du1ft,3du1full} and to divide the whole 
investigation into two steps. The finite-temperature model is formulated on 
anisotropic lattice with different spatial and temporal couplings. Here, 
following~\cite{3du1ft}, as a first step, we study the limit of 
vanishing spatial coupling. 
Since this approximation does not affect the global symmetry properties of the 
model, we hope that in this limit the model belongs to the same universality 
class as the full theory. It is well known that for both three- and 
four-dimensional gauge models at finite temperature this is indeed the case, 
at least for $N=2,3,4$. Furthermore, in this limit the gauge model can be 
mapped onto a generalized $Z(N)$ spin model in $2d$. Thus, this investigation 
sheds some light on the details of the critical behavior of the general Potts 
model, {\it i.e.} beyond the vector Potts model, which is usually 
considered in the literature. In the present paper we study the phase 
transitions in models with $N=5,7$ in great details and in addition we locate 
critical points of the models with $N=9,13$. Our computations are performed 
on lattices with temporal extent $N_t=2,4$ and with spatial size in the 
range $L\in [128-2048]$. 

This paper is organized as follows. In Section~2 we formulate our model and 
study some of its properties analytically. In particular, we establish the
exact relation with a generalized $2d$ $Z(N)$ spin model and discuss some of 
the RG predictions regarding the critical behavior. Also, we give simple 
analytical estimates for the critical couplings. In Section~3 we present
the setup of Monte Carlo simulations, define the observables used in this work
and present the numerical results of simulations for $N=5,7$. In 
Section~4 we present results for the critical points in the models with 
$N=9,13$ and discuss the scaling of the critical points with $N$. Our 
conclusions and perspectives are given in Section~5.   

\section{Analytical considerations}

\subsection{Relation of the $3d$ $Z(N)$ LGT to a generalized $2d$ $Z(N)$ vector model}

We work on a $3d$ lattice $\Lambda = L^2\times N_t$ with spatial extension $L$ 
and temporal extension $N_t$; $\vec{x}=(x_0,x_1,x_2)$, where $x_0\in [0,N_t-1]$
and $x_1,x_2\in [0,L-1]$ denote the sites of the lattice
and $e_n$, $n=0,1,2$, denotes a unit vector in the $n$-th direction.
Periodic boundary conditions (BC) on gauge fields are imposed in all 
directions. The notations $p_t$ ($p_s$) stand for the temporal (spatial) 
plaquettes, $l_t$ ($l_s$) for the temporal (spatial) links.

We introduce conventional plaquette angles $s(p)$ as
\begin{equation}
s(p) \ = \ s_n(x) + s_m(x+e_n) - s_n(x+e_m) - s_m(x) \ .
\label{plaqangle}
\end{equation}
The $3d$ $Z(N)$ gauge theory on an anisotropic lattice can generally be 
defined as 
\begin{equation}
Z(\Lambda ;\beta_t,\beta_s;N) \ = \  \prod_{l\in \Lambda}
\left ( \frac{1}{N} \sum_{s(l)=0}^{N-1} \right ) \ \prod_{p_s} Q(s(p_s)) \
\prod_{p_t} Q(s(p_t)) \ .
\label{PTdef}
\end{equation}
The most general $Z(N)$-invariant Boltzmann weight with $N-1$ different 
couplings is
\begin{equation}
Q(s) \ = \
\exp \left [ \sum_{k=1}^{N-1} \beta_p(k) \cos\frac{2\pi k}{N}s \right ] \ .
\label{Qpgen}
\end{equation}
The Wilson action corresponds to the choice $\beta_p(1)=\beta_p$, 
$\beta_p(k)=0, k=2,...,N-1$. The $U(1)$ gauge model is defined as the limit 
$N\to\infty$ of the above expressions.

To study the phase structure of $3d$ $Z(N)$ LGTs in the strong coupling 
limit ($\beta_s = 0$) one can map the gauge model to a generalized $2d$ spin 
$Z(N)$ model with the action 
\begin{equation}
\label{modaction}
S \ =\ \sum_{x}\ \sum_{n=1}^2 \sum_{k = 1}^{N-1} \ \beta_k \  
\cos \left( \frac{2 \pi k}{N} \left(s(x) - s(x+e_n) \right) \right) \ .
\end{equation}
The effective coupling constants $\beta_k$ are derived from the coupling 
constant $\beta_t \equiv \beta$ of the $Z(N)$ LGT, using the following 
equation (the Wilson action is used for the gauge model):
\begin{equation}
\beta_k \ =\ \frac{1}{N} \sum_{p = 0}^{N - 1} \ln(Q_p) \cos \left(\frac{2 \pi 
p k}{N} \right) \ ,
\label{couplings}
\end{equation}
where
\begin{eqnarray}
\label{Qk}
Q_k&\ =\ &\sum_{p = 0}^{N - 1} \ \left(\frac{B_p}{B_0}\right)^{N_t}  \ 
\cos \left(\frac{2 \pi p k}{N} \right) \ ,   \\
B_k&\ =\ &\sum_{p = 0}^{N - 1} \exp \left[ \beta \cos \left(\frac{2 \pi p}{N} 
\right) \right ] 
\cos \left(\frac{2 \pi p k}{N} \right) \ .
\label{couplings_coeff}
\end{eqnarray}
These equations can be easily obtained in a few steps following similar 
computations for $3D$ $SU(N)$ model in the same limit $\beta_s=0$ (see, 
for example, Ref.~\cite{eff_action} and references therein): 
\begin{itemize}
\item
Fourier expansion of the original Boltzmann weight;
\item
Integration over spatial gauge fields; this leads to an effective $2d$ model 
for the Polyakov loops with the Boltzmann weight $Q(s(x)-s(x+e_n))$ defined 
in~(\ref{Qk});
\item
Exponentiation and re-expansion in a new Fourier series.
\end{itemize}

\subsection{Renormalization group and critical behavior} 

To gain some information on the critical behavior of the gauge model at 
$\beta_s=0$, one can perform the RG study of the theory. Let us consider the 
$2d$ model obtained after integration over spatial gauge fields,
\begin{equation} 
Z(\Lambda ;\beta; N) \ = \ \prod_{x\in \Lambda} 
\left ( \frac{1}{N} \sum_{s(x)=0}^{N-1} \right ) \ \prod_{l} \ 
\left [  \sum_{k = 0}^{N - 1} \ \left(B_k \right )^{N_t}  \ 
\cos \left(\frac{2 \pi k}{N} (s(x) - s(x+e_n)) \right)  \right ] \ .
\label{PF_orig}
\end{equation}
The coefficients $B_k\equiv B_k(\beta)$ are given in~(\ref{couplings_coeff})
and can be represented as 
\begin{equation} 
B_k \ = \ \sum_{r=-\infty}^{\infty} \ I_{Nr+k}(\beta)  \ .
\label{Bk_exp}
\end{equation}
Here, $I_k(x)$ is the modified Bessel function. The spin variables $s(x)$ 
can be associated with the Polyakov loops of the original model. 

The RG equations can be obtained only for the Villain formulation of the model.
Replacing the Bessel function with its asymptotics and using the Poisson 
summation formula, one finds
\begin{eqnarray} 
Z(\Lambda ;\beta; N) \ &=& \ \prod_{x\in \Lambda} 
\left ( \frac{1}{N} \sum_{s(x)=0}^{N-1} \right ) \\
&\times& \prod_{l} \left [  \sum_{m=-\infty}^{\infty}
\exp \left [ -\frac{1}{2}\beta_{\rm eff} 
\left ( \frac{2\pi }{N}(s(x) - s(x+e_n))+2\pi m \right )^2 \right ]  \right ] 
\nonumber \ .
\label{PF_villain}
\end{eqnarray}
This is nothing but the Villain formulation of $2d$ $Z(N)$ vector model with 
$\beta_{\rm eff}$ defined as 
\begin{equation} 
\beta_{\rm eff} \ = \ \beta/N_t \ .
\label{b_eff}
\end{equation}
The RG equations for the model~(\ref{PF_villain}) have been constructed 
in~\cite{elitzur}. Their analysis has been performed by us in a recent 
work~\cite{2dzn7_17}. 
Therefore, all the conclusions as to the critical behavior remain valid in the 
present case. 
We shortly list the main results below:
\begin{enumerate}
\item The critical RG trajectories in the planes $(\beta_{\rm eff},y)$ and 
$(\beta_{\rm eff},z)$, where $y$, $z$ are the activities of the vortex 
configurations, coincide with those of the $2d$ $Z(N)$ vector spin model;

\item The critical index $\nu$ has been computed numerically from the 
RG equations for a variety of $N$. It takes on the value 1/2 for all $N$ 
considered, in particular for $N=5,7$;

\item The calculation of the two-point correlation function reveals that 
the index $\eta$ is equal to 1/4 at the first transition point, while 
$\eta=4/N^2$ at the second critical point~\cite{elitzur}. 
The same behavior is valid for the correlation of the Polyakov loops in our 
model; 

\item In~\cite{2dzn7_17} we have calculated the dependence of the critical 
points on $N$. These scaling formulae are expected to hold in the present case 
for any fixed $N_t$ and will be the subject of the discussion in Section~4.  

\end{enumerate}

\subsection{Estimation of the critical points}

The location of the critical points in $3d$ $Z(N)$ LGTs for $N>4$ 
is unknown. Therefore, before presenting numerical results, it is instructive 
to give some simple analytical predictions for the critical values of 
$\beta^{\rm crit}$ for different values $N_t$. Such a prediction could serve 
then as the starting point for the numerical search of the critical points. 
Following~\cite{3du1ft}, such critical values can be easily estimated if one 
knows $\beta^{\rm crit}$ for $N_t=1$. Since the model with $N_t=1$ coincides 
with the $2d$ $Z(N)$ model, approximate critical points for other values of 
$N_t$ can be computed from the equality
\begin{equation}
\frac{B_1}{B_0}(\beta^{\rm crit}) \ = 
\ \left [ \frac{B_1}{B_0}(x)\right]^{N_t} \ ,
\label{crpointest}
\end{equation}
where $x$ on the right-hand side denotes the unknown critical point. 
Solving the last equation numerically one finds $x$. As for the critical 
points $\beta^{\rm crit}$ at $N_t=1$, we use our previous estimates 
from~\cite{2dzn}. 
As it will be seen below, the predicted values are in a reasonable agreement 
with the numerical results. Table~\ref{tbl:bpc_pred} summarizes approximate 
analytical predictions for the critical couplings at the two transitions,
denoted by $\beta_{\rm c}^{(1)}$ and $\beta_{\rm c}^{(2)}$, respectively; the 
last two columns show Monte Carlo values. 

\begin{table}[h]
\caption{Values of $\beta_{\rm c}^{(1)}$ and $\beta_{\rm c}^{(2)}$ expected 
for $N_t = 1,\ 2,\ 4$ in $Z(N)$ with $N = 5,\ 7$.}
\begin{center}
\begin{tabular}{|c|c|c|c|c|c|}
\hline
$N$ & $N_t$ & $\beta_{\rm c}^{(1)}$ & $\beta_{\rm c}^{(2)}$ & 
${\beta_{\rm c}^{(1)}}_{\rm MC}$ & ${\beta_{\rm c}^{(2)}}_{\rm MC}$\\
\hline
 5 & 1 & - & - & 1.051(1) & 1.105(1) \\
 5 & 2 & 1.8393 & 1.9057 & 1.87(1) & 1.940(7) \\
 5 & 4 & 2.7761 & 2.8515 & 2.813(3) & 2.898(4) \\
\hline
 7 & 1 & - & - & 1.111(1) & 1.88(8) \\
 7 & 2 & 1.9861 & 3.1303 & 2.031(7) & 3.366(7) \\
 7 & 4 & 3.2995 & 4.9669 & 3.406(8) & 5.158(7) \\
\hline
\end{tabular}
\end{center}
\label{tbl:bpc_pred}
\end{table}

\section{Numerical results}

\subsection{Setup of the Monte Carlo simulation}

To study the phase transitions we used the cluster algorithm described 
in~\cite{2dzn}. The model is studied on a square $L \times L$ lattice 
$\Lambda$ with periodic BC. Simulations were carried on for $N_t = 2,\ 4$,
but could easily be done also for other values of $N_t$, since the 
parameter $N_t$ appears only in the definition of the 
couplings~(\ref{couplings}). 
As original action of the gauge model we used the conventional Wilson action.
For each Monte Carlo run the typical number of generated configurations 
was $10^6$, the first $5\times10^4$ of them being discarded to ensure 
thermalization. Measurements were taken after 10 updatings and 
the uncertainty on primary observables was estimated by the jackknife 
method combined with binning.

We considered the following observables:
\begin{itemize}
\item complex magnetization $M_L = |M_L| e^{i \psi}$,
\begin{equation}
\label{complex_magnetization}
M_L \ =\  \sum_{x \in \Lambda} \exp \left( \frac{2 \pi i}{N} s(x) \right) \;;
\end{equation}

\item population $S_L$,
\begin{equation}
\label{population}
S_L \ =\  \frac{N}{N - 1} \left(\frac{\max_{i = 0, N - 1} n_i} {L^2} 
- \frac{1}{N} \right)\;,  
\end{equation}
where $n_i$ is number of $s(x)$ equal to $i$;

\item real part of the rotated magnetization $M_R = |M_L| \cos(N \psi)$
and normalized rotated magnetization $m_\psi = \cos(N \psi)$;

\item susceptibilities of $M_L$, $S_L$ and $M_R$:  
$\chi_L^{(M)}$, $\chi_L^{(S)}$, $\chi_L^{(M_R)}$,
\begin{equation}
\label{susceptibilities}
\chi_L^{(\mathbf\cdot)} \ =\  L^2 \left(\left< \mathbf\cdot^2 \right> 
- \left< \mathbf\cdot \right>^2 \right)\;;
\end{equation}

\item Binder cumulants $U_L^{(M)}$ and $B_4^{(M_R)}$,
\begin{eqnarray}
U_L^{(M)}&\ =\ &1 - \frac{\left\langle \left| M_L \right| ^ 4 
\right\rangle}{3 \left\langle \left| M_L \right| ^ 2 \right\rangle^2}\;, 
\nonumber \\
\label{binderU}
B_4^{(M_R)}&\ =\ & \frac{\left\langle \left| M_R 
- \left\langle M_R \right\rangle \right| ^ 4 \right\rangle}
{\left\langle \left| M_R - \left\langle M_R \right\rangle \right| ^ 2 
\right\rangle ^ 2 }\;;
\label{binderBMR}
\end{eqnarray}

\item helicity modulus $\Upsilon$,
\begin{equation}
\Upsilon \ =\  \left \langle e \right \rangle - L^2 \beta \left \langle s^2 
\right \rangle \ ,
\label{helicity_modulus}
\end{equation}
where   
\begin{equation*}
e \equiv \frac{1}{L^{2}} \sum_{ { \left \langle ij \right \rangle }_{x}} 
\cos{\left(\theta_{i}-\theta_{j}
\right)} \ ,\ s\equiv\frac{1}{L^{2}} \sum_{ {\left \langle ij \right 
\rangle}_{x}} \sin{(\theta_{i}
-\theta_{j})}\ ,\ \theta_i \equiv \frac{2\pi}{N} s(i)
\end{equation*} 
and the notation $ {\left \langle ij \right \rangle}_{x} $ means 
nearest-neighbors spins in the $x$-direction.
\end{itemize}

\subsection{Determination of the critical couplings}

\begin{figure}
\centering
\includegraphics[width=0.32\textwidth]{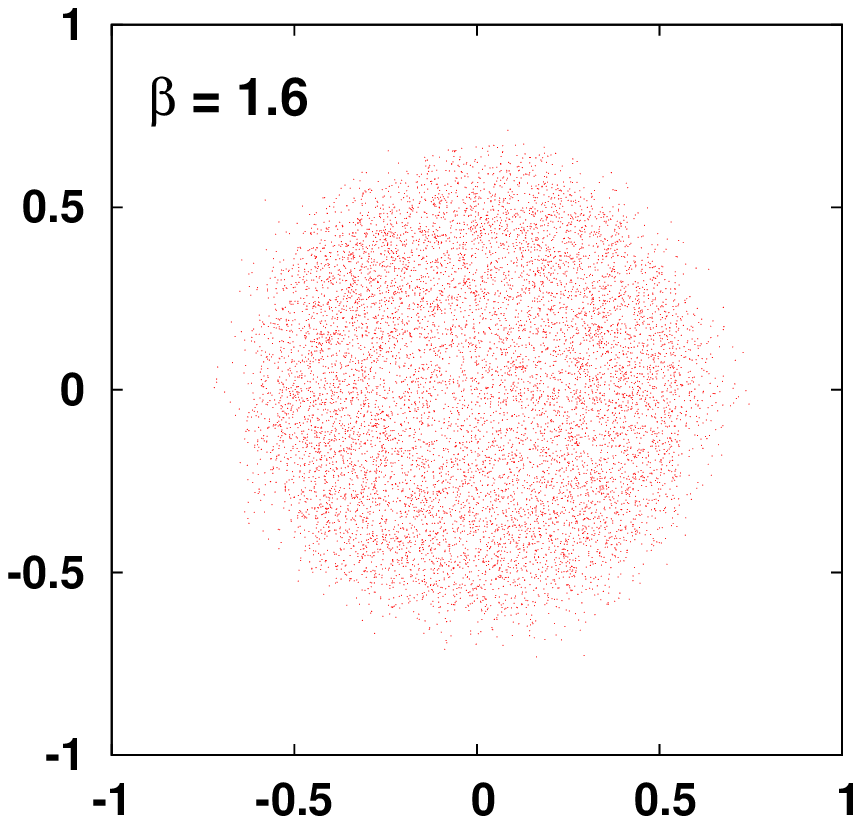}
\includegraphics[width=0.32\textwidth]{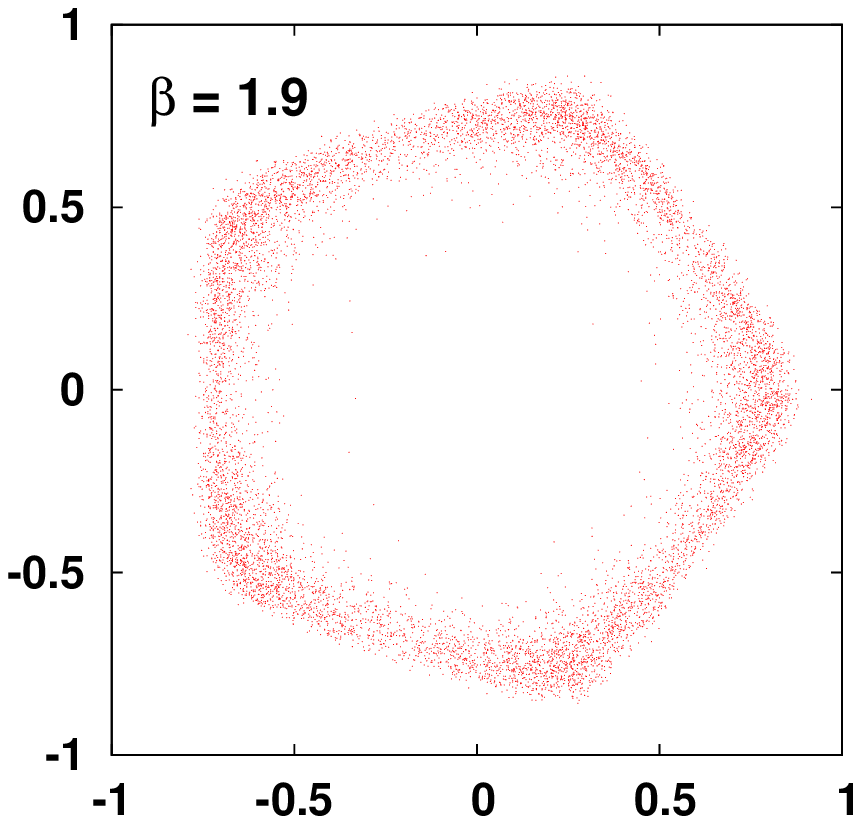}
\includegraphics[width=0.32\textwidth]{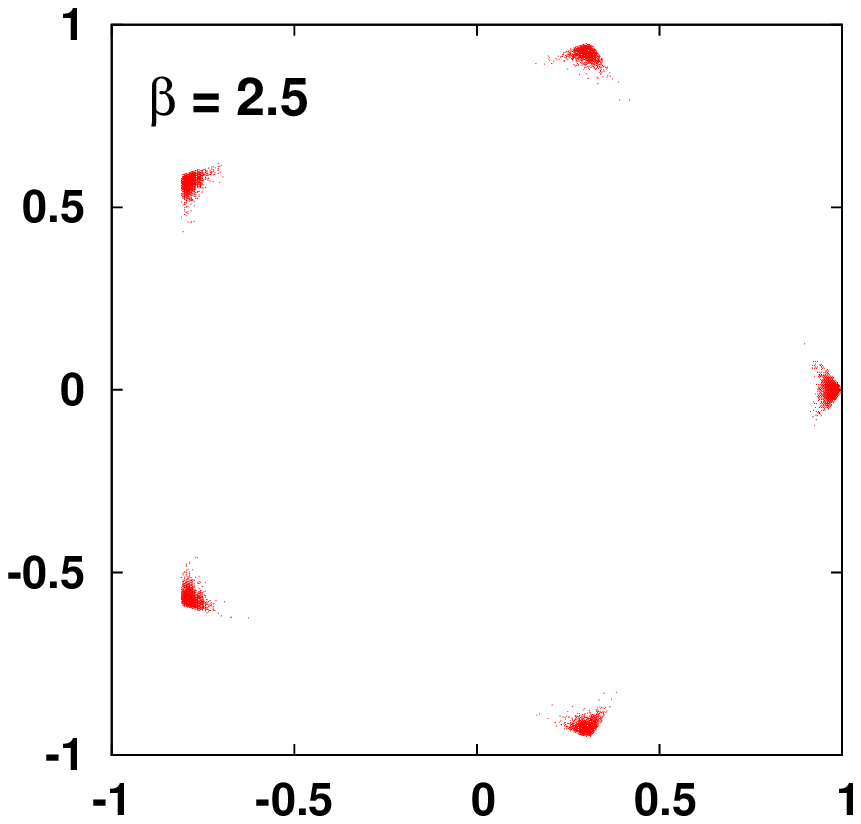}
\caption{Scatter plot of the complex magnetization $M_L$ at $\beta$=1.6, 1.9 
and 2.5 in $Z(5)$ on a $16^2\times 2$ lattice.}
\label{fig:scatter}
\end{figure}

A clear indication of the three-phase structure emerges from the inspection
of the scatter plot of the complex magnetization $M_L$ at different values 
of $\beta$: as we move from low to high $\beta$, we observe the transition
from a disordered phase (uniform distribution around zero) through an
intermediate phase (ring distribution) up to the ordered 
phase ($N$ isolated spots), as Fig.~\ref{fig:scatter} shows for the case of 
$Z(5)$ on a $16^2\times 2$ lattice.

The first and most important numerical task is to determine the value
of the two critical couplings in the thermodynamic limit, $\beta_c^{(1)}$ 
and $\beta_c^{(2)}$, that separate the three phases. To this aim we have 
adopted several methods, which we list here:
\begin{itemize}
\item Methods for the determination of $\beta_{\rm c}^{(1)}$: 

(a) locate the position $\beta_{\rm pc}^{(1)}(L)$ of the peak of the 
susceptibility $\chi_L^{(M)}$ of the complex magnetization $|M_L|$ on lattices
with various spatial size and find $\beta_{\rm c}^{(1)}$ by a fit with 
the following scaling function, dictated by the essential scaling:
\begin{equation}
\beta^{(1)}_{\rm pc}=\beta^{(1)}_{\rm c}
+\frac{A}{(\ln L + B)^{\frac{1}{\nu}}}\quad ,
\label{b_pc}
\end{equation}
taking $\nu$ equal to 1/2;

(b) estimate the crossing point of the curves giving the behavior of the
Binder cumulant $U_L^{(M)}$ versus $\beta$ on lattices with different
spatial size $L$ or, alternatively, search for the value of 
$\beta_{\rm c}^{(1)}$ which optimizes the overlap of these curves when they 
are plotted against $(\beta-\beta_{\rm c}^{(1)})(\ln L)^{1/\nu}$, with $\nu$ 
fixed at 1/2; 

(c) consider the helicity modulus $\Upsilon$ near the phase transition and 
define $\beta_{\rm pc}^{(1)}(L)$ as the value of $\beta$ such that
$\eta(\beta) \equiv 1/(2 \pi \beta \Upsilon)=1/4$
on the lattice with spatial size $L$~\cite{YO91}, then find $\beta_{\rm c}^{(1)}$ by a 
fit with the scaling function 
\begin{equation}
\beta^{(1)}_{\rm pc}=\beta^{(1)}_{\rm c} + \frac{A}{\ln L + B}\;,
\label{helicity_scaling}
\end{equation}
valid under the assumption that the phase transition belongs to the $XY$ 
universality class.

\item Methods for the determination of $\beta_{\rm c}^{(2)}$: 

(d) same as the method (a) using instead the susceptibility $\chi_L^{(S)}$ 
of the population $S_L$;

(e) same as the method (b) using instead simultaneously the Binder cumulant $B_4^{(M_R)}$
and the order parameter $m_\psi$. 

\end{itemize}

\begin{figure}
\includegraphics[width=0.45\textwidth]{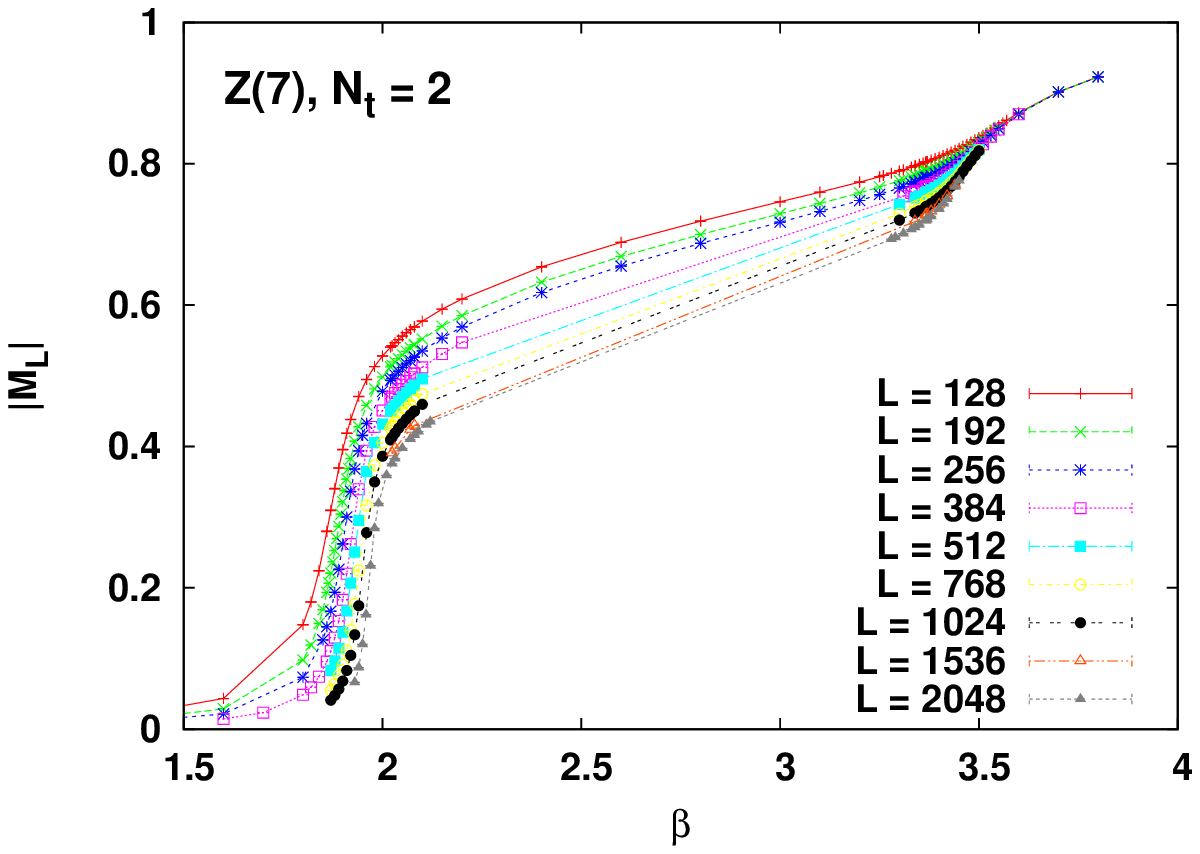}
\includegraphics[width=0.45\textwidth]{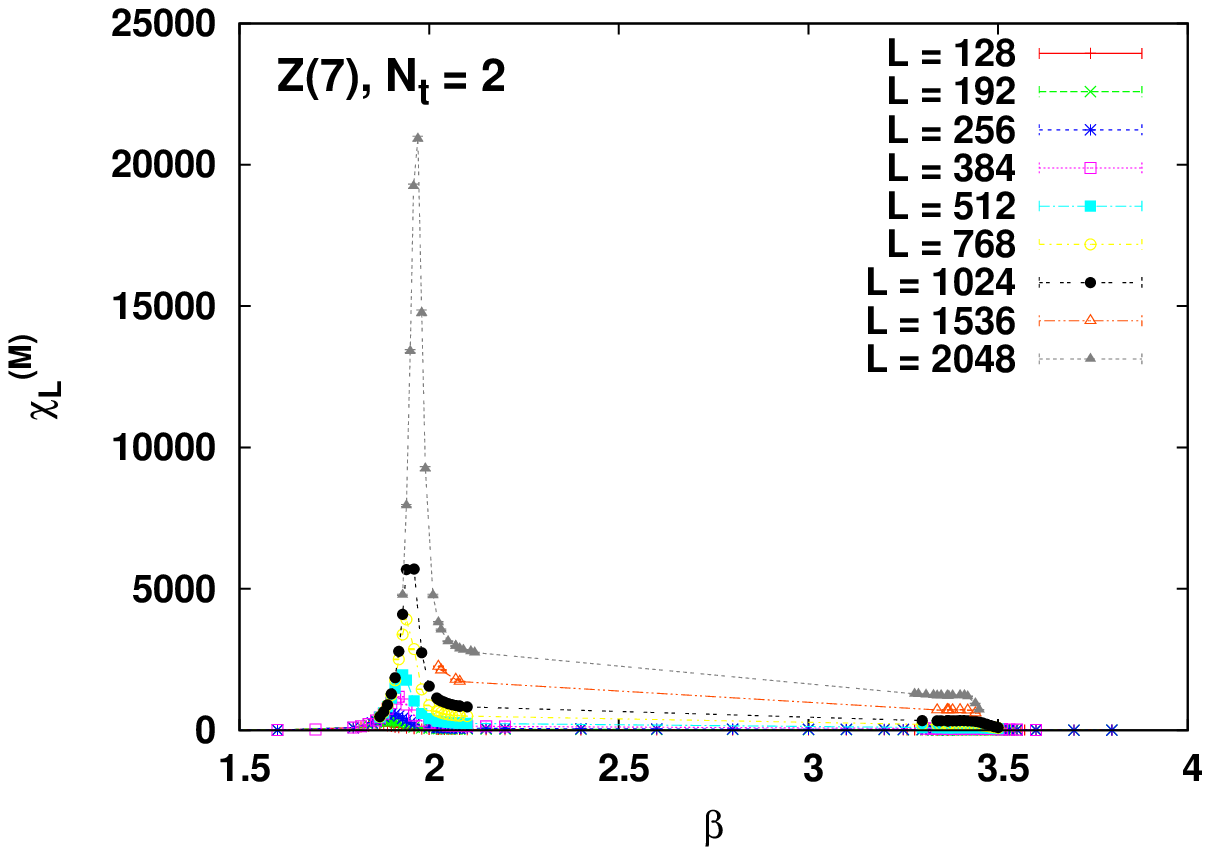}
\caption{Behavior of $M_L$ and of its susceptibility versus $\beta$ in 
$Z(7)$ on lattices with $N_t=2$ and $L$ ranging from 128 to 2048.}
\label{chi_ML_Z7_Nt2}
\end{figure}

\begin{figure}
\includegraphics[width=0.45\textwidth]{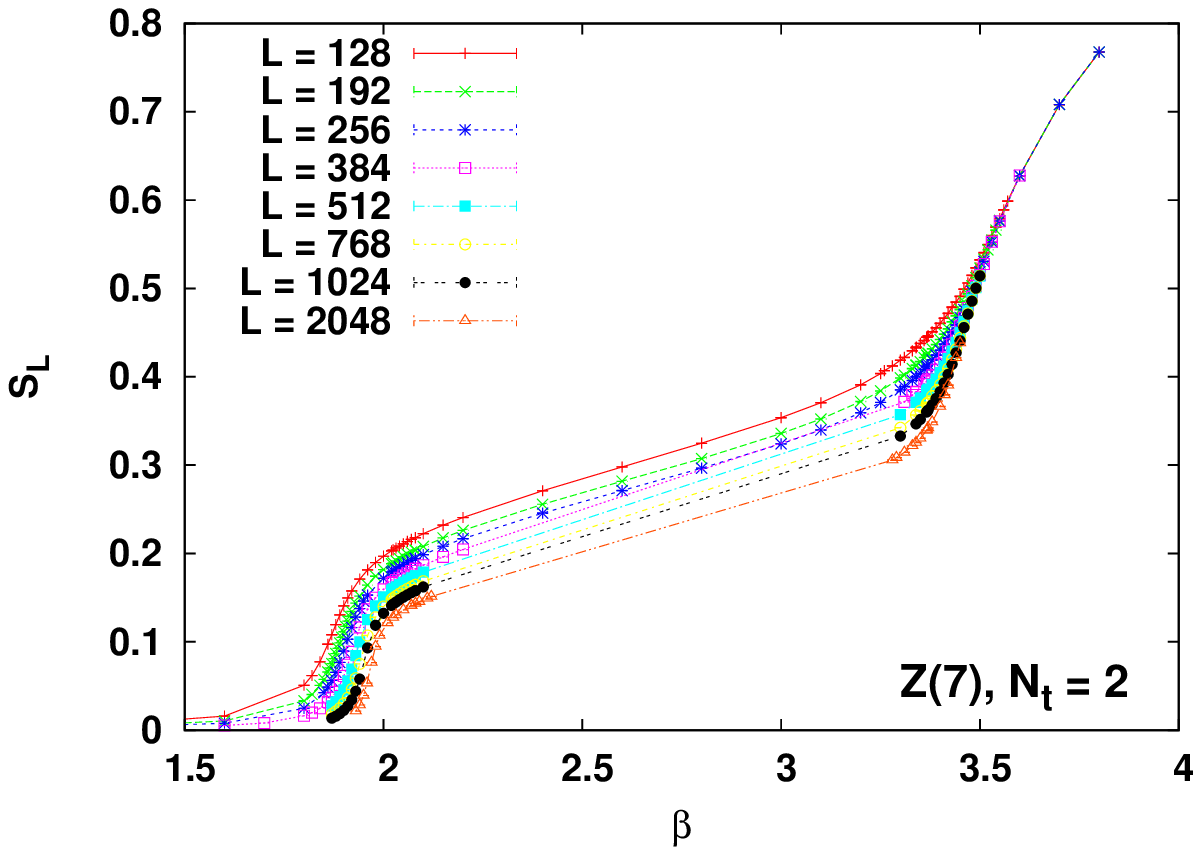}
\includegraphics[width=0.45\textwidth]{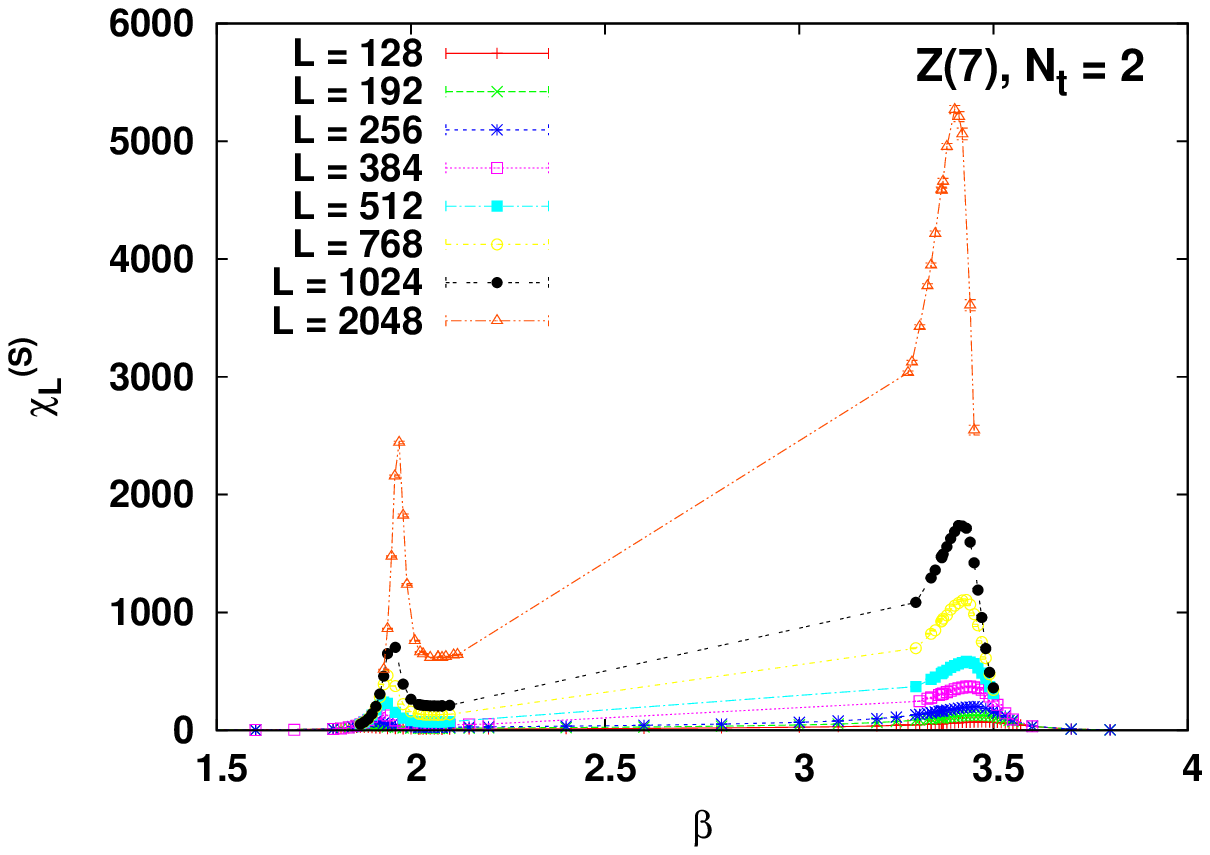}
\caption{Behavior of $S_L$ and of its susceptibility versus $\beta$ in 
$Z(7)$ on lattices with $N_t=2$ and $L$ ranging from 128 to 2048.}
\label{chi_SL_Z7_Nt2}
\end{figure}

As an illustration of the methods (a) and (d), we show in 
Figs.~\ref{chi_ML_Z7_Nt2} and~\ref{chi_SL_Z7_Nt2} the behavior of $M_L$ (and 
of its susceptibility) and that of $S_L$ (and of its susceptibility) versus 
$\beta$ in $Z(7)$ on lattices with $N_t=2$ and $L$ ranging from 128 to 2048.
In Table~\ref{peaks} we summarize all the values of $\beta_{\rm pc}^{(1)}(L)$
and $\beta_{\rm pc}^{(2)}(L)$ found in this work for the application of 
methods (a) and (d) in $Z(N)$ with $N=5,7$ for $N_t=2,4$.

\begin{table}
\caption{Summary of all the determinations of $\beta_{\rm pc}^{(1)}$
and $\beta_{\rm pc}^{(2)}$ in $Z(N)$ on lattices with size 
$L^2\times N_t$.}
\begin{center}
\begin{tabular}{|c|c|r|c|c|}
\hline
$N$ & $N_t$ & $L$ & $\beta_{\rm pc}^{(1)}$ & $\beta_{\rm pc}^{(2)}$ \\
\hline
    &       & 128 & 1.7656(1)  & 1.9773(5) \\
    &       & 192 & 1.7808(1)  & 1.9740(3) \\
    &       & 256 & 1.7910(1)  & 1.9713(4) \\
 5  &   2   & 384 & 1.80124(7) & 1.9689(5) \\
    &       & 512 & 1.80774(6) & 1.9658(3) \\
    &       & 768 & 1.81540(5) & 1.9626(2) \\
    &       &1024 & 1.82013(4) & 1.9614(4) \\
    &       &2048 & 1.83001(5) &           \\
\hline
\hline
    &       &  16 & 2.4913(9)  & \\
    &       &  32 & 2.5928(6)  & \\
    &       &  64 & 2.6538(5)  & \\
    &       & 128 & 2.692(1)   & 2.9376(8) \\
 5  &   4   & 192 & 2.7131(8)  & 2.934(2)  \\
    &       & 256 & 2.7226(7)  & 2.928(1)  \\
    &       & 384 & 2.7357(7)  & 2.927(2)  \\
    &       & 512 &            & 2.921(2)  \\
    &       & 768 &            & 2.920(3)  \\
    &       &1024 &            & 2.917(2)  \\
\hline
\hline
    &       & 128 & 1.8644(4)  & 3.461(2)  \\
    &       & 192 & 1.8873(2)  & 3.454(1)  \\
    &       & 256 & 1.9011(2)  & 3.443(2)  \\
 7  &   2   & 384 & 1.9184(2)  & 3.433(1)  \\
    &       & 512 & 1.9289(2)  & 3.427(2)  \\
    &       & 768 & 1.9421(2)  & 3.420(2)  \\
    &       &1024 & 1.9504(5)  & 3.416(2)  \\
\hline
\hline
    &       & 128 & 3.14(1)    & 5.257(1)  \\
    &       & 192 & 3.18(1)    & 5.242(2)  \\
    &       & 256 & 3.20(1)    & 5.2371(7) \\
 7  &   4   & 384 & 3.2220(2)  & 5.223(2)  \\
    &       & 512 & 3.2392(1)  & 5.216(2)  \\
    &       & 768 & 3.26024(6) & 5.209(2)  \\
    &       &1024 & 3.2737(3)  & 5.198(2)  \\
\hline
\end{tabular}
\end{center}
\label{peaks}
\end{table}

\begin{figure}
\includegraphics[width=0.45\textwidth]{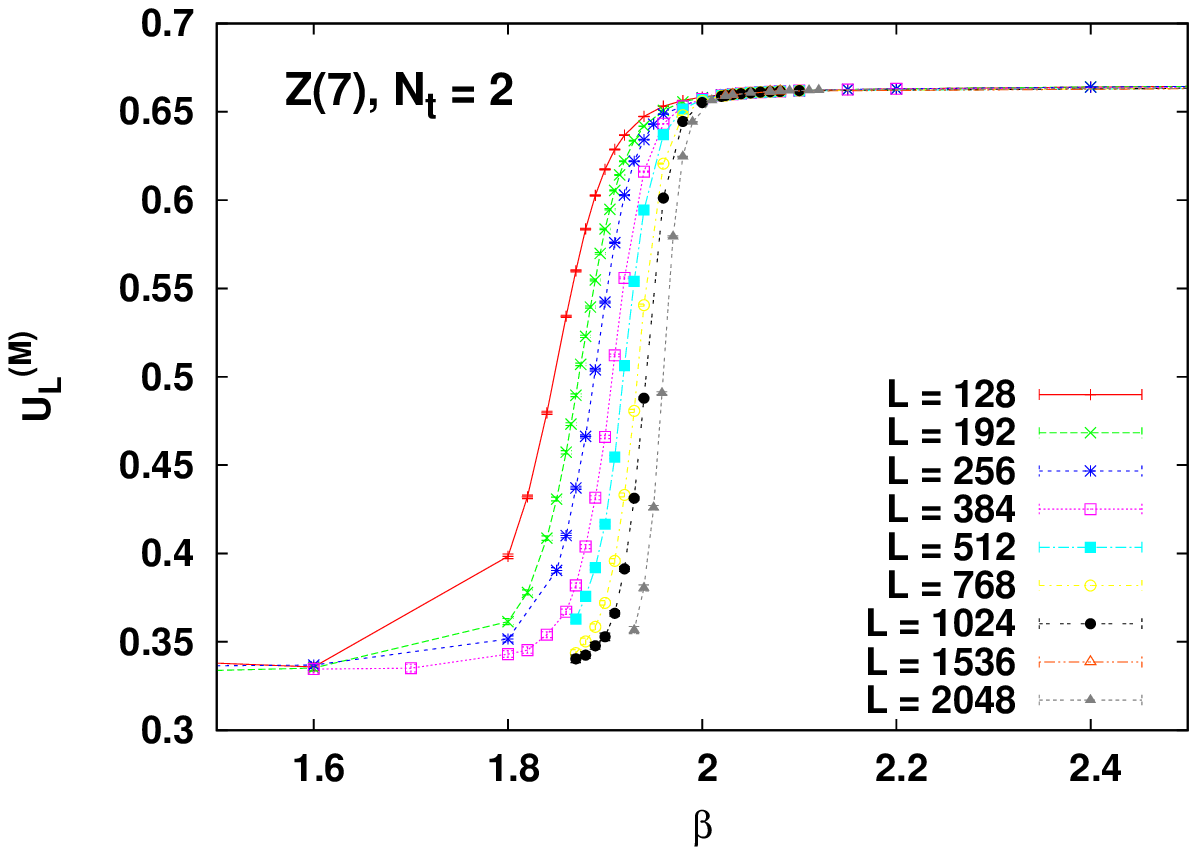}
\includegraphics[width=0.45\textwidth]{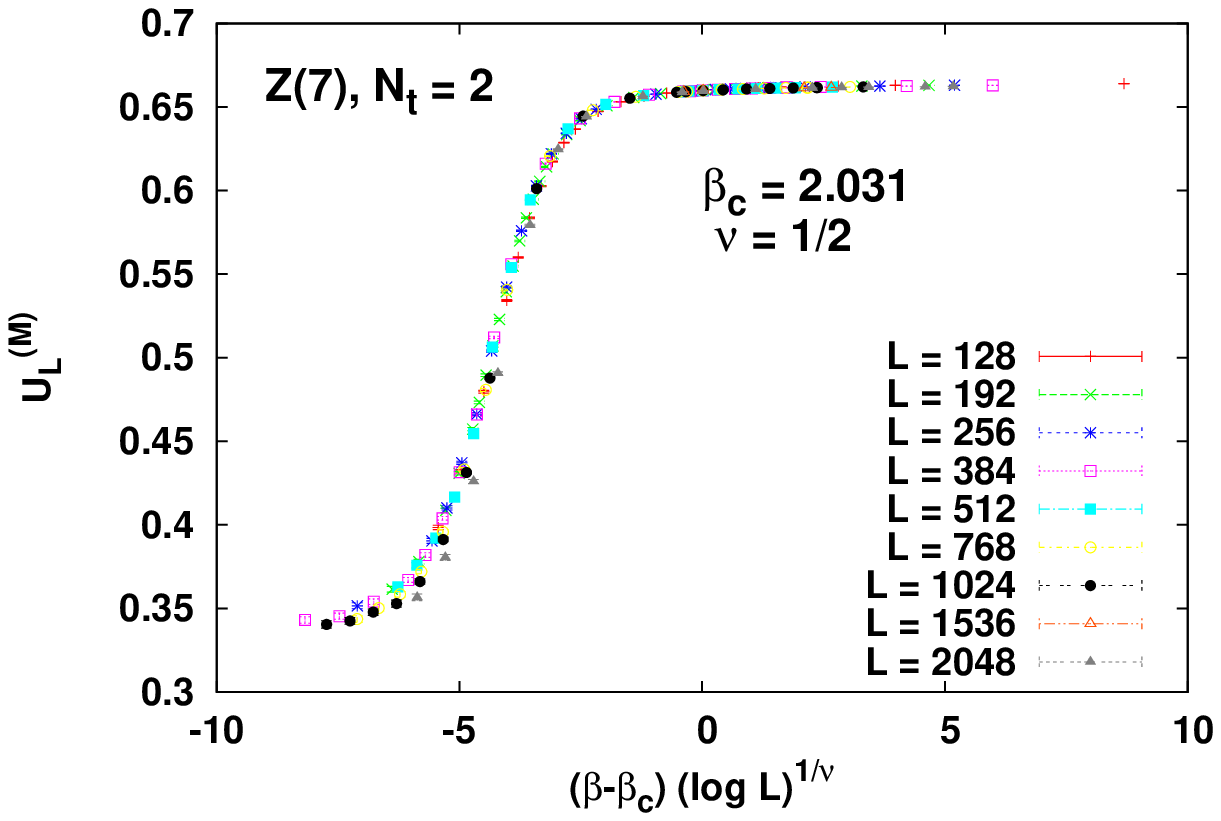}
\caption{Binder cumulant $U_L^{(M)}$ as function of $\beta$ (left) and
of $(\beta-\beta_c) (\ln L)^{1/\nu}$ (right) in $Z(7)$ on lattices with 
$N_t=2$ and $L$ ranging from 128 to 2048.}
\label{UL_Z7_Nt2}
\end{figure}

\begin{figure}
\includegraphics[width=0.45\textwidth]{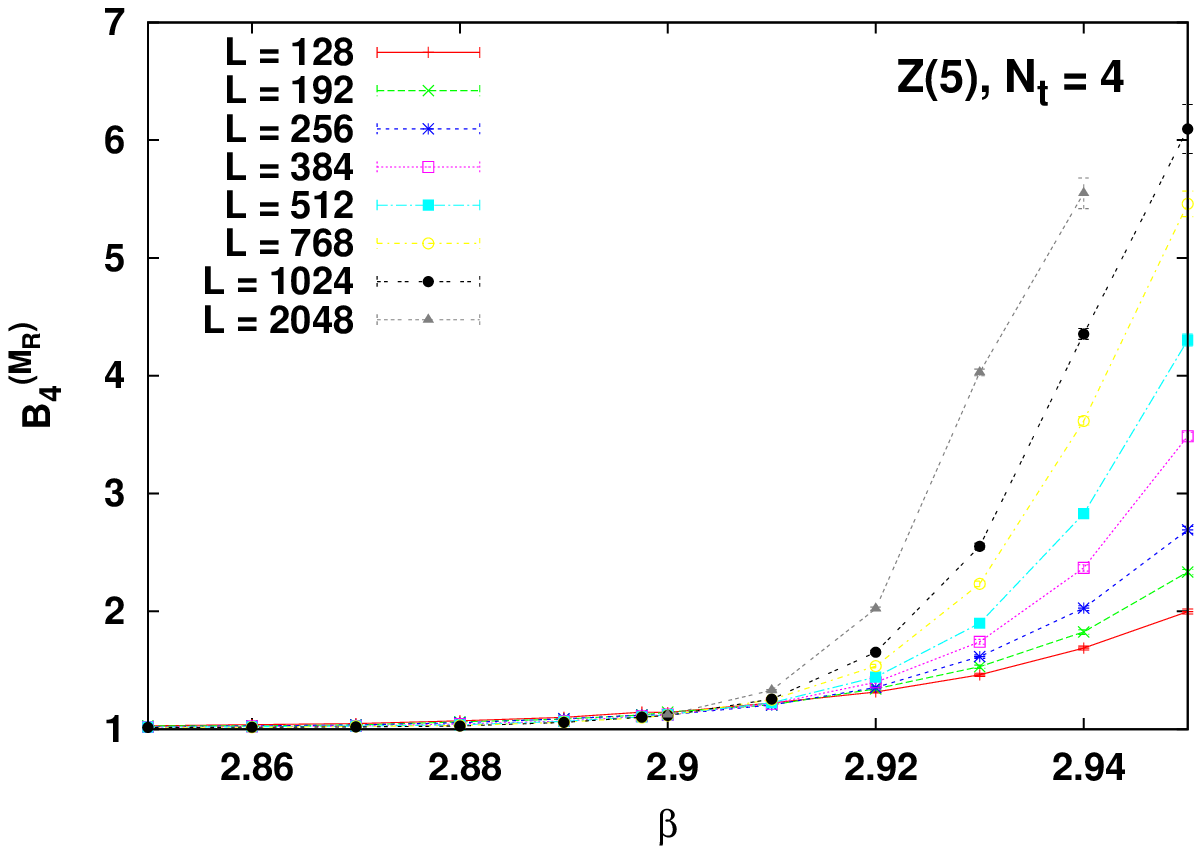}
\includegraphics[width=0.45\textwidth]{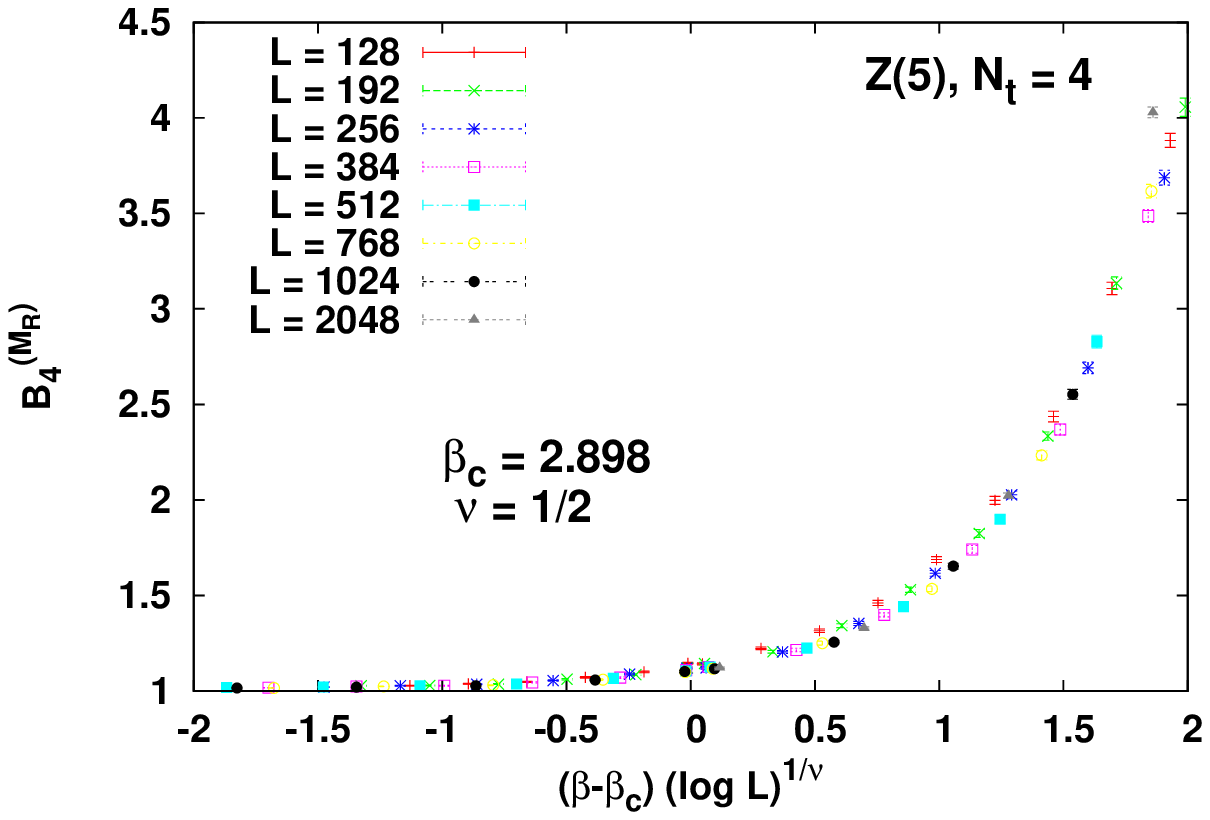}
\caption{Binder cumulant $B_4^{(M_R)}$ as a function of $\beta$ (left)
and of $(\beta-\beta_{\rm c}) (\ln L)^{1/\nu}$ (right) in $Z(5)$ on
lattices with $N_t=4$ and $L$ ranging from 128 to 2048.}
\label{B4_Z5_Nt4}
\end{figure}

\begin{figure}
\includegraphics[width=0.45\textwidth]{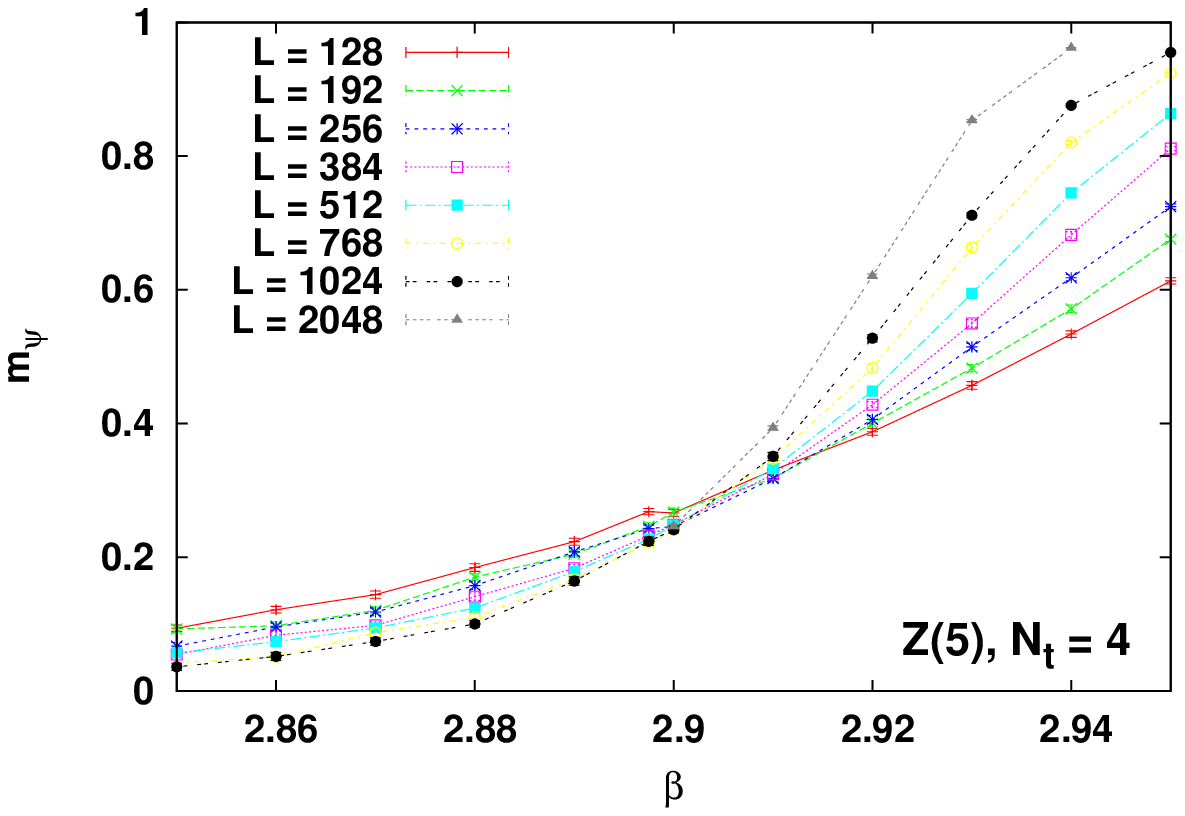}
\includegraphics[width=0.45\textwidth]{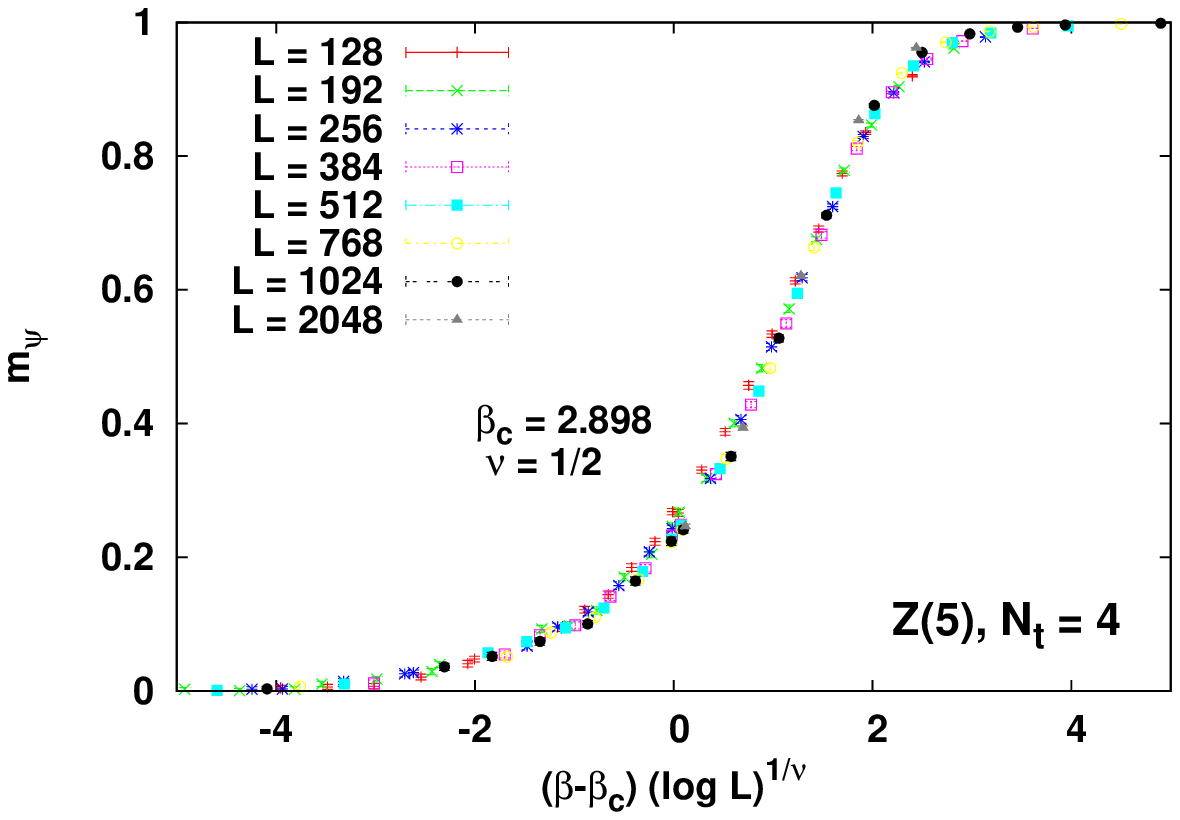}
\caption{$m_{\psi}$ as a function of $\beta$ (left) and
of $(\beta-\beta_{\rm c}) (\ln L)^{1/\nu}$ (right) in $Z(5)$ on
lattices with $N_t=4$ and $L$ ranging from 128 to 2048.}
\label{mpsi_Z5_Nt4}
\end{figure}

As an illustration of the method (b), we show in Fig.~\ref{UL_Z7_Nt2}
the behavior of $U^{(M)}_L$ versus $\beta$ in $Z(7)$ on lattices with 
$N_t=2$ and $L$ ranging from 128 to 2048. Similarly, as an illustration
of the method (e), we show in Figs.~\ref{B4_Z5_Nt4} 
and~\ref{mpsi_Z5_Nt4} the behavior of $B_4^{(M_R)}$ and of
$m_\psi$ versus $\beta$ in $Z(5)$ on lattices with $N_t=4$ and $L$ ranging 
from 128 to 2048.

\begin{figure}
\includegraphics[width=0.45\textwidth]{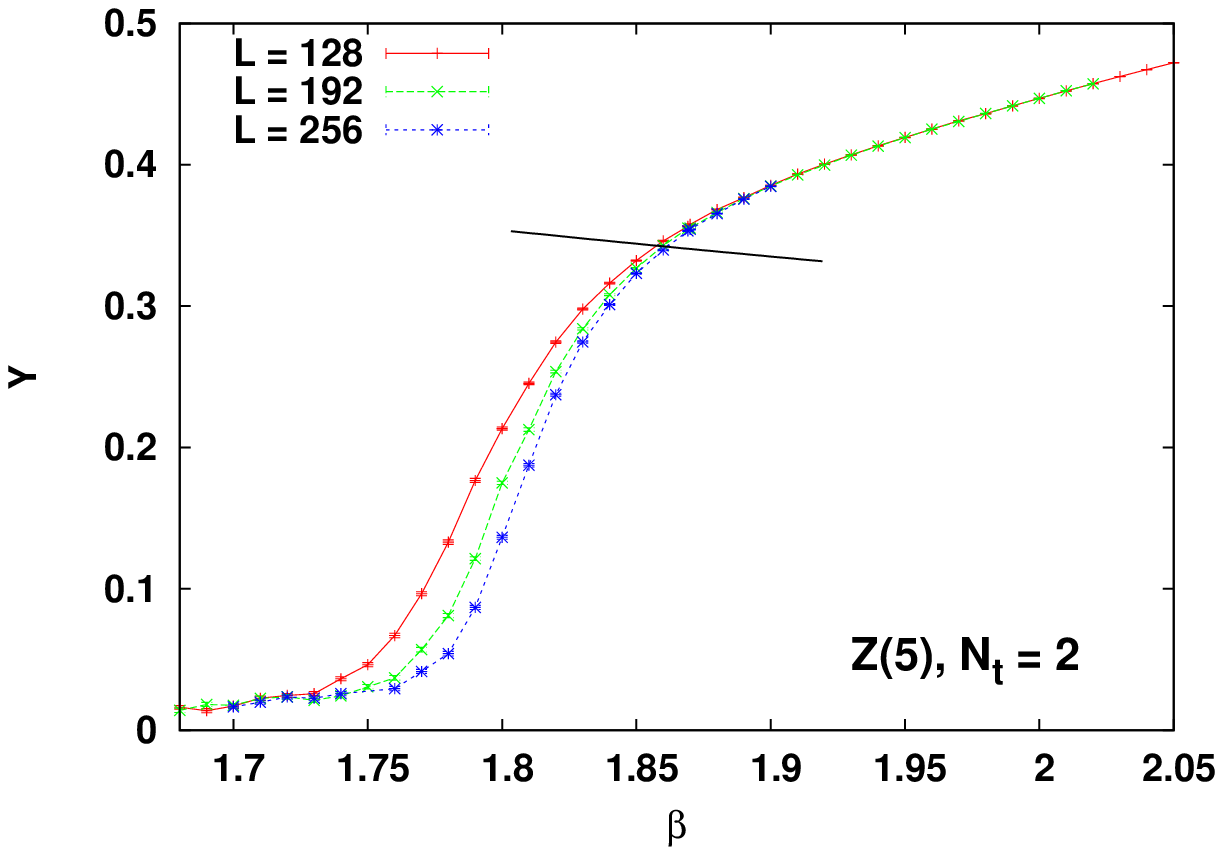}
\includegraphics[width=0.45\textwidth]{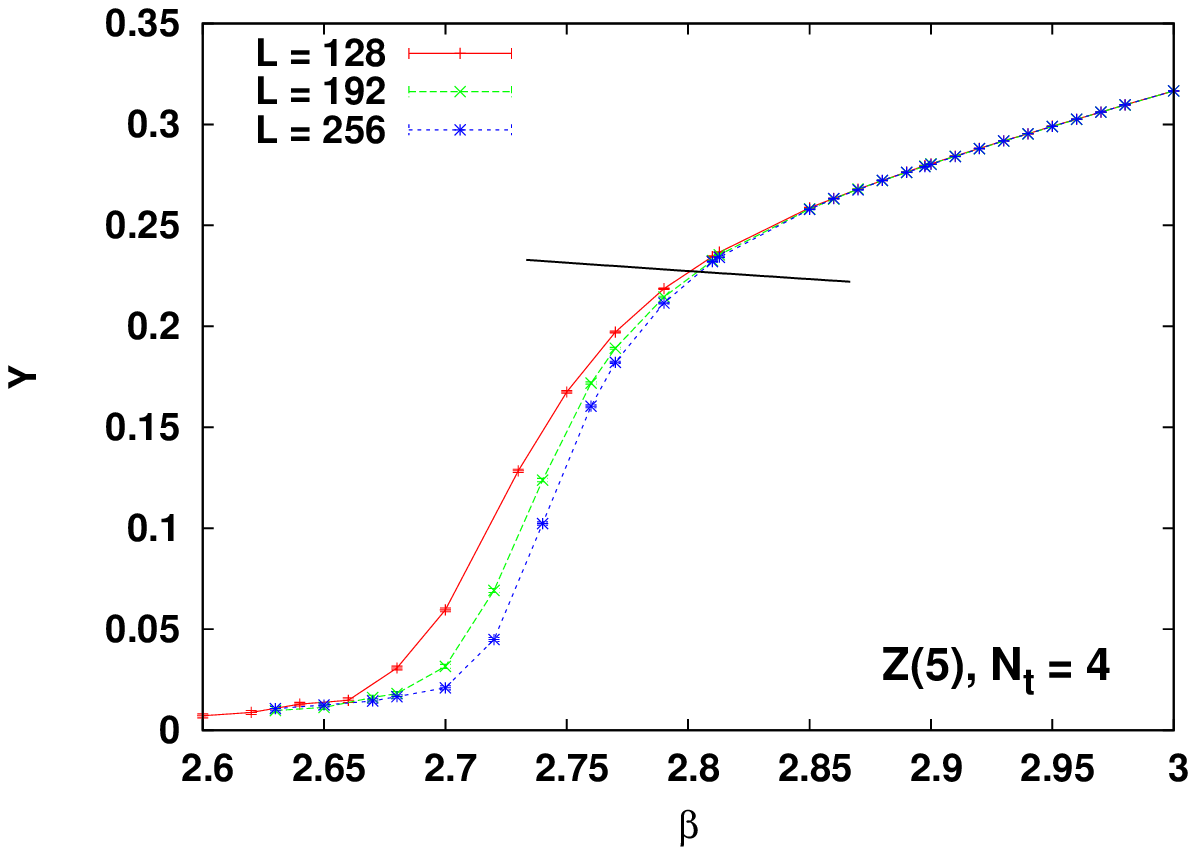}

\includegraphics[width=0.45\textwidth]{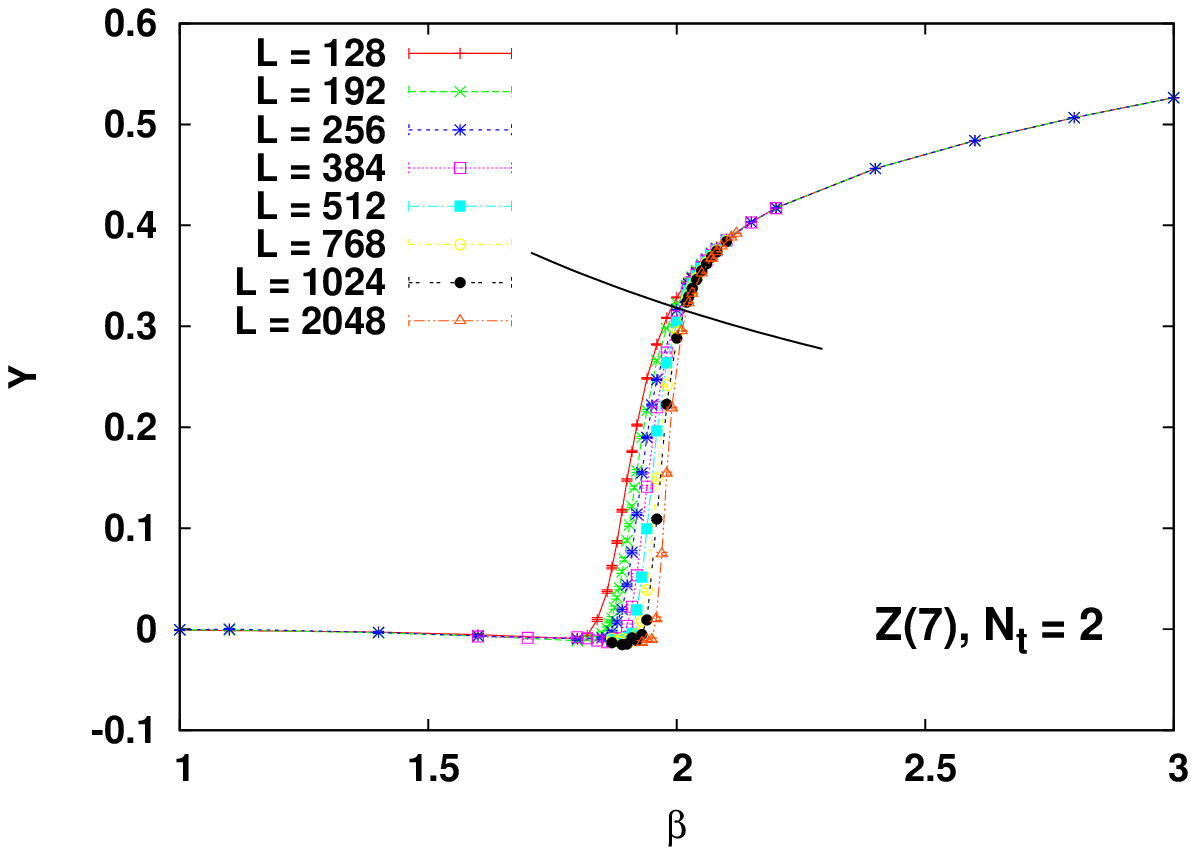}
\includegraphics[width=0.45\textwidth]{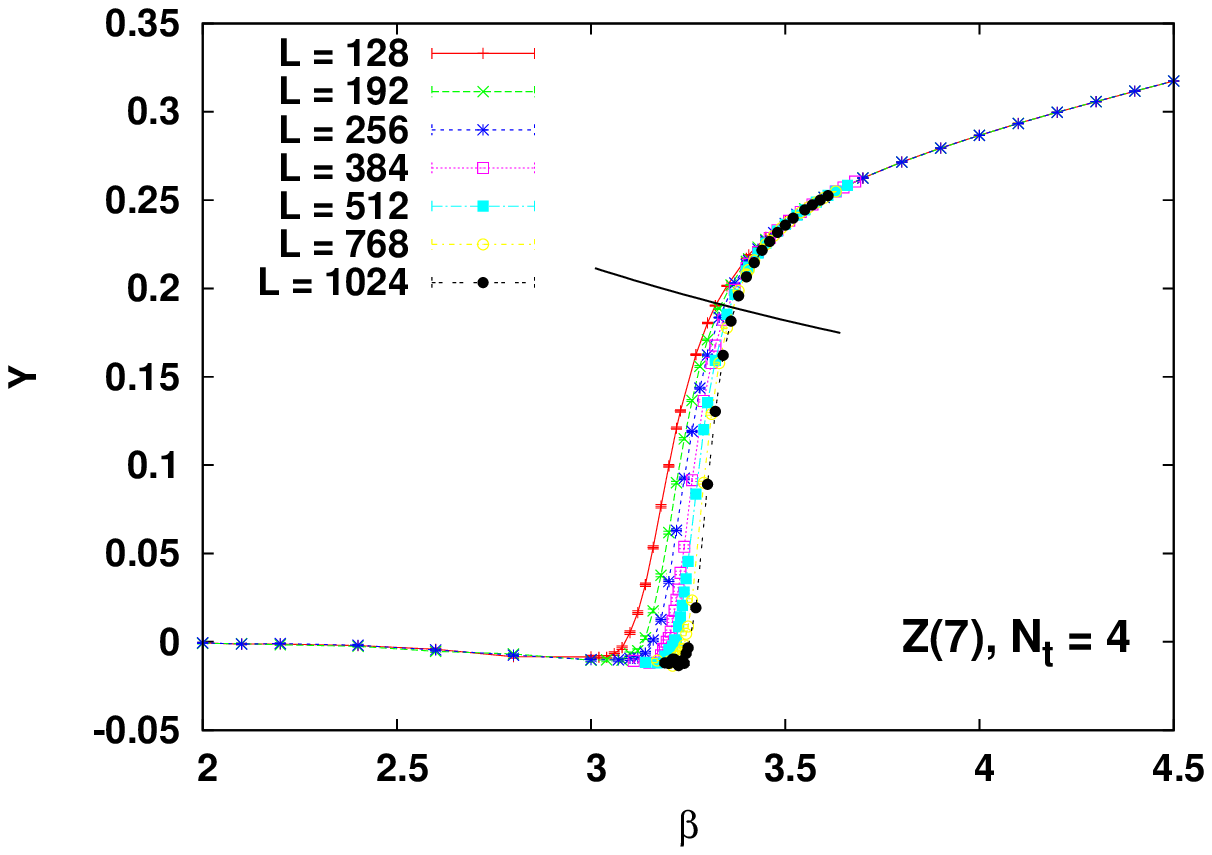}
\caption{Helicity modulus as a function of $\beta$ in $Z(5)$ with $N_t=2$
(top left), in $Z(5)$ with $N_t=4$ (top right), in $Z(7)$ with $N_t=2$ 
(bottom left) and in $Z(7)$ with $N_t=4$ (bottom right).} 
\label{helicity}
\end{figure}

As an illustration of the method (c), we show in Fig.~\ref{helicity}
the behavior of the helicity modulus $\Upsilon$ versus $\beta$ along with the 
line $\Upsilon = 1/(2\pi\beta\eta),\ \eta = 1/4$, describing pseudocritical points, 
in $Z(5)$ and in $Z(7)$ on lattices with $N_t=2,4$ and several values of the
spatial extension $L$. For larger values of $L$ only the points 
around the intersection were simulated.

Finally, we report in Table~\ref{crit_betas} the determinations of the
critical couplings $\beta_{\rm c}^{(1)}$ and $\beta_{\rm c}^{(2)}$
in $Z(N)$ with $N$=5 and 7 for $N_t$=2 and 4, specifying the adopted method.

\begin{table}
\caption{Summary of the determinations of $\beta_{\rm c}^{(1)}$ and
and $\beta_{\rm c}^{(2)}$ in $Z(N)$ on lattices with size $L^2\times N_t$.
The fourth (sixth) column gives the method adopted to find 
$\beta_{\rm c}^{(1)}$ ($\beta_{\rm c}^{(2)}$).}
\begin{center}
\begin{tabular}{|c|c|c|c|c|c|}
\hline
$N$ & $N_t$ & $\beta_{\rm c}^{(1)}$ & method & $\beta_{\rm c}^{(2)}$ & method\\
\hline
    &       & 1.878(2)              &   a    & 1.91(6)              &   d    \\
  5 &   2   & 1.87(1)               &   b    & 1.940(7)             &   e    \\
    &       & 1.8801(7)             &   c    &                      &        \\
\hline
    &       & 2.832(3)              &   a    & 2.88(3)              &   d    \\
  5 &   4   & 2.813(3)              &   b    & 2.898(4)             &   e    \\
    &       & 2.829(1)              &   c    &                      &        \\
\hline
    &       & 2.070(2)              &   a    & 3.31(9)              &   d    \\
  7 &   2   & 2.031(7)              &   b    & 3.366(7)             &   e    \\
    &       & 2.069(1)              &   c    &                      &        \\
\hline
    &       & 3.47(1)               &   a    & 5.0(2)               &   d    \\
  7 &   4   & 3.406(8)              &   b    & 5.158(7)             &   e    \\
    &       & 3.4782(8)             &   c    &                      &        \\
\hline
\end{tabular}
\end{center}
\label{crit_betas}
\end{table}

\subsection{Determination of critical indices at the two transitions}
\label{sec:betas}

Once critical couplings have been estimated, we are able to extract some
critical indices and check the hyperscaling relation. 

We start the discussion from the first transition. According to the 
standard finite-size scaling (FSS) theory, the equilibrium magnetization 
$|M_{L}|$ at criticality should obey the relation 
$|M_{L}| \sim L^{-\beta / \nu}$, if the spatial extension $L$ of the lattice
is large enough~\footnote{The symbol $\beta$ here denotes a critical index and 
not, obviously, the coupling of the theory. In spite of this inconvenient 
notation, we are confident that no confusion will arise, since it will be 
always clear from the context which $\beta$ is to be referred to.}. Therefore,
we fit data of $|M_L|$ at $\beta^{(1)}_{\rm c}$ on all lattices with size 
$L$ not smaller than a given $L_{\rm min}$ with the scaling law
\begin{equation}
|M_{L}|=A L^{-\beta/\nu} \ln^r L\;,
\label{magn_fss}
\end{equation}
where a non-zero value for $r$ takes into account the possibility of 
logarithmic corrections~\cite{Kenna-Irving,Hasenbusch}.

The FSS behavior of the susceptibility $\chi^{(M)}_L$ is given by 
$\chi^{(M)}_L\sim L^{\gamma/ \nu}$, where $\gamma/\nu=2-\eta$ and $\eta$ is 
the magnetic critical index. Therefore we fit data of $\chi^{(M)}_L$ at
$\beta^{(1)}_{\rm c}$ on all lattices with size $L$ not smaller 
than a given $L_{\rm min}$ according to the scaling law
\begin{equation}
\chi^{(M)}_{L}= A L^{\gamma/\nu} \ln^r L\;.
\label{chiM_fss}
\end{equation}
As the value of the critical coupling $\beta^{(1)}_{\rm c}$ we use the
central value of the determination from the method (b) (see 
Table~\ref{crit_betas}). 

The results of the fits are summarized in 
Tables~\ref{indices_Z5_Nt2},~\ref{indices_Z5_Nt4},~\ref{indices_Z7_Nt2},~\ref{indices_Z7_Nt4}, for the cases of $Z(5)$ with $N_t$=2, $Z(5)$ with $N_t$=4,
$Z(7)$ with $N_t$=2 and $Z(7)$ with $N_t$=4, respectively. The reference 
value for the index $\eta$ at this transition is 1/4, whereas the      
hyperscaling relation to be fulfilled is $\gamma/\nu+2\beta/\nu=d=2$.
 
\begin{table}
\caption{Critical indices $\beta/\nu$ and $\gamma/\nu$ for the first 
transition in $Z(5)$ with $N_t$ = 2, determined by the fits given in 
Eqs.~(\ref{magn_fss}) and~(\ref{chiM_fss}) on the complex magnetization
$M_L$ and its susceptibility $\chi_L^{(M)}$ at $\beta_{\rm c}^{(1)}= 1.87$ 
for different choices of the minimum lattice size $L_{\rm min}$
(an asterisk indicates a fixed parameter). The $\chi^2$ of the two fits, 
given in the columns four and seven, is the reduced one.}
\begin{center}
\setlength{\tabcolsep}{2.5pt}
\begin{tabular}{|c|c|c|c|c|c|c|c|c|}
\hline
$L_{\min}$ & $\beta/\nu$  & $r_{\beta/\nu}$  & $\chi^2_{\beta/\nu}$ 
           & $\gamma/\nu$ & $r_{\gamma/\nu}$ & $\chi^2_{\gamma/\nu}$ 
           & $d = 2\beta/\nu + \gamma/\nu$   & $\eta = 2-\gamma/\nu$ \\
\hline
  16 & 0.12040(7) & 0$^\ast$  & 4.43 & 1.692(2) & 0$^\ast$     & 52.74 & 1.933(2) & 0.308(2) \\ 
     & 0.1228(5)  & 0.013(3)  & 2.43 & 1.97(1)  & $-$1.44(6)   & 0.90  & 2.22(1)  & 0.03(1) \\ 
     &            &           &      & 1.668(2) & 0.125$^\ast$ & 61.75 & 1.908(2) & 0.332(2) \\ 
\hline
  32 & 0.12047(8) & 0$^\ast$  & 4.49 & 1.712(2) & 0$^\ast$     & 22.49 & 1.953(2) & 0.288(2) \\ 
     &  0.1239(7) &  0.019(4) & 2.12 &  1.97(2) & $-$1.4(1)    & 0.99  &  2.22(2) &  0.03(2) \\ 
     &            &          &       & 1.690(2) & 0.125$^\ast$ & 26.40 & 1.931(2) & 0.310(2) \\ 
\hline
  64 &  0.1206(1) & 0$^\ast$ & 4.30  & 1.729(3) & 0$^\ast$     & 6.93  & 1.970(3) & 0.271(3) \\ 
     &   0.126(1) & 0.034(7) & 1.51 &  1.93(3)  & $-$1.2(2)    & 0.73  &  2.18(3) &  0.07(3) \\ 
     &            &          &      & 1.708(3)  & 0.125$^\ast$ & 8.30  & 1.949(3) & 0.292(3) \\ 
\hline
 128 &  0.1208(1) & 0$^\ast$ & 2.43 & 1.740(3)  & 0$^\ast$     & 3.00  & 1.982(3) & 0.260(3) \\ 
     &   0.126(2) &  0.03(1) & 1.72 &  1.94(9)  & $-$1.3(5)    & 0.84  &  2.19(9) &  0.06(9) \\ 
     &            &          &      & 1.720(3)  & 0.125$^\ast$ & 3.48  & 1.962(3) & 0.280(3) \\ 
\hline
 192 &  0.1210(1) & 0$^\ast$ & 2.09 & 1.745(4)  & 0$^\ast$     & 1.97  & 1.987(4) & 0.255(4) \\ 
     &   0.125(3) &  0.03(2) & 2.06 &  1.91(4)  & $-$1.1(2)    & 0.97  &  2.16(4) &  0.09(4) \\ 
     &            &          &      & 1.726(4)  & 0.125$^\ast$ & 2.25  & 1.968(4) & 0.274(4) \\ 
\hline
 256 &  0.1211(2) & 0$^\ast$ & 2.19 & 1.752(5)  & 0$^\ast$     & 1.44  & 1.994(5) & 0.248(5) \\ 
     &   0.124(4) &  0.02(3) & 2.56 & 1.906(5)  & $-$1.02(3)   & 1.19  & 2.15(1) & 0.094(5) \\ 
     &            &          &      & 1.733(5)  & 0.125$^\ast$ & 1.56  & 1.975(5) & 0.267(5) \\ 
\hline
 384 &  0.1212(2) & 0$^\ast$ & 2.44 & 1.758(6)  & 0$^\ast$     & 1.02  & 2.000(6) & 0.242(6) \\ 
     &  0.1224(2) & 0.0079(9)& 3.27 & 1.836(6)  & $-$0.53(3)   & 1.37  & 2.080(6) & 0.164(6) \\ 
     &            &          &      & 1.740(6)  & 0.125$^\ast$ & 1.04  & 1.982(6) & 0.260(6) \\ 
\hline
 512 &  0.1215(3) & 0$^\ast$ & 2.18 & 1.762(7)  & 0$^\ast$     & 1.06  & 2.005(8) & 0.238(7) \\ 
     &  0.1021(2) & $-$0.135(1)& 1.21 & 1.840(6)  &$-$0.54(3)  & 1.88  & 2.044(7) & 0.160(6) \\ 
     &            &          &      & 1.744(7)  & 0.125$^\ast$ & 1.02  & 1.987(8) & 0.256(7) \\ 
\hline
 768 &  0.1209(4) & 0$^\ast$ & 1.71 &  1.76(1)  & 0$^\ast$     & 1.49  & 2.00(1) &  0.24(1) \\ 
     &  0.1220(3) &  0.008(1)& 3.52 & 1.834(8)  & $-$0.54(3)   & 3.24  & 2.078(8) & 0.166(8) \\ 
     &            &          &      &  1.74(1)  & 0.125$^\ast$ & 1.47  &  1.98(1) &  0.26(1) \\ 
\hline
1024 &  0.1209(6) & 0$^\ast$ & 3.41 &  1.76(2)  & 0$^\ast$     & 2.93  &  2.00(2) &  0.24(2) \\ 
     &            &          &      &  1.74(2)  & 0.125$^\ast$ & 2.89  &  1.98(2) &  0.26(2) \\ 
\hline
\end{tabular}
\end{center}
\label{indices_Z5_Nt2}
\end{table}

\begin{table}
\caption{The same as Table~\ref{indices_Z5_Nt2} for $Z(5)$ with $N_t=4$,
determined at $\beta_c^{(1)} = 2.813$.}
\begin{center}
\setlength{\tabcolsep}{2.5pt}
\begin{tabular}{|c|c|c|c|c|c|c|c|c|}
\hline
$L_{\min}$ & $\beta/\nu$  & $r_{\beta/\nu}$  & $\chi^2_{\beta/\nu}$ 
           & $\gamma/\nu$ & $r_{\gamma/\nu}$ & $\chi^2_{\gamma/\nu}$ 
           & $d = 2 \beta/\nu + \gamma/\nu$  & $\eta = 2-\gamma/\nu$\\
\hline
 128 & 0.1215(2) & 0$^\ast$ & 0.45 & 1.730(5) & 0$^\ast$     & 1.23 & 1.973(5) & 0.270(5) \\ 
     &  0.126(4) &  0.02(2) & 0.30 & 1.808(6) & $-$0.46(3)   & 1.36 &  2.06(1) & 0.192(6) \\ 
     &           &          &      & 1.709(5) & 0.125$^\ast$ & 1.32 & 1.952(5) & 0.291(5) \\ 
\hline
 192 & 0.1217(3) & 0$^\ast$ & 0.25 & 1.734(6) & 0$^\ast$     & 1.20 & 1.977(7) & 0.266(6) \\ 
     & 0.1228(3) & 0.007(1) & 0.32 & 1.707(7) &  0.17(3)     & 1.59 & 1.952(7) & 0.293(7) \\ 
     &           &          & 0.25 & 1.714(6) & 0.125$^\ast$ & 1.19 & 1.957(7) & 0.286(6) \\ 
\hline
 256 & 0.1218(3) & 0$^\ast$ & 0.30 & 1.740(8) & 0$^\ast$     & 1.07 & 1.983(8) & 0.260(8) \\ 
     & 0.1229(3) & 0.007(1) & 0.47 &  1.62(3) &   0.7(2)     & 0.88 &  1.87(3) &  0.38(3) \\ 
     &           &          &      & 1.720(8) & 0.125$^\ast$ & 0.98 & 1.963(8) & 0.280(8) \\ 
\hline
 384 & 0.1217(5) & 0$^\ast$ & 0.44 &  1.73(1) & 0$^\ast$     & 0.24 &  1.97(1) &  0.27(1) \\ 
     & 0.1228(4) & 0.007(2) & 0.86 & 1.785(9) & $-$0.38(4)   & 0.62 &  2.03(1) & 0.215(9) \\ 
     &           &          &      &  1.71(1) & 0.125$^\ast$ & 0.22 &  1.95(1) &  0.29(1) \\ 
\hline
 512 & 0.1220(7) & 0$^\ast$ & 0.49 &  1.72(2) & 0$^\ast$     & 0.05 &  1.96(2) &  0.28(2) \\ 
     &           &          &      &  1.70(2) & 0.125$^\ast$  & 0.05 &  1.94(2) &  0.30(2) \\ 
\hline
\end{tabular}
\end{center}
\label{indices_Z5_Nt4}
\end{table}

\begin{table}
\caption{The same as Table~\ref{indices_Z5_Nt2} for $Z(7)$ with $N_t=2$,
determined at $\beta_c^{(1)} = 2.031$.}
\begin{center}
\setlength{\tabcolsep}{2.5pt}
\begin{tabular}{|c|c|c|c|c|c|c|c|c|}
\hline
$L_{\min}$ & $\beta/\nu$  & $r_{\beta/\nu}$  & $\chi^2_{\beta/\nu}$ 
           & $\gamma/\nu$ & $r_{\gamma/\nu}$ & $\chi^2_{\gamma/\nu}$ 
           & $d = 2 \beta/\nu + \gamma/\nu$  & $\eta = 2- \gamma/\nu$\\
\hline
 128 & 0.12862(9) & 0$^\ast$  & 11.95 & 1.768(2) & 0$^\ast$     & 0.93 & 2.025(2) & 0.232(2) \\ 
     &   0.141(1) &  0.078(9) & 0.93  & 1.763(2) & 0.030(3)     & 1.15 & 2.045(5) & 0.237(2) \\ 
     &            &           &       & 1.747(2) & 0.125$^\ast$ & 1.18 & 2.004(2) & 0.253(2) \\ 
\hline
 192 &  0.1290(1) & 0$^\ast$  & 7.73 & 1.767(3)  & 0$^\ast$     & 1.09 & 2.025(3) & 0.233(3) \\ 
     &   0.143(2) &   0.09(1) & 0.87 &  1.89(2)  & $-$0.8(1)    & 0.33 &  2.17(3) &  0.11(2) \\ 
     &            &           &      & 1.748(3)  & 0.125$^\ast$ & 1.36 & 2.006(3) & 0.252(3) \\ 
\hline
 256 &  0.1294(1) & 0$^\ast$  & 5.84 & 1.771(4)  & 0$^\ast$     & 0.83 & 2.029(4) & 0.229(4) \\ 
     &   0.145(3) &   0.10(2) & 0.95 &  1.87(3) & $-$0.6(2)     & 0.44 &  2.16(3) &  0.13(3) \\ 
     &            &           &      & 1.752(4) & 0.125$^\ast$  & 1.00 & 2.010(4) & 0.248(4) \\ 
\hline
 384 &  0.1297(2) & 0$^\ast$  & 4.96 & 1.775(5) & 0$^\ast$      & 0.58 & 2.034(5) & 0.225(5) \\ 
     &  0.1517(2) & 0.1480(8) & 0.27 & 1.865(5) & $-$0.61(2)    & 0.55 & 2.168(5) & 0.135(5) \\ 
     &            &           &      & 1.757(5) & 0.125$^\ast$  & 0.64 & 2.016(5) & 0.243(5) \\ 
\hline
 512 &  0.1303(2) & 0$^\ast$  & 2.43 & 1.775(6) & 0$^\ast$      & 0.77 & 2.035(7) & 0.225(6) \\ 
     &  0.1504(2) & 0.1389(9) & 0.38 &  1.94(6) &  $-$1.2(4)      & 0.32 &  2.25(6) &  0.06(6) \\ 
     &            &           &      & 1.757(6) & 0.125$^\ast$  & 0.86 & 2.017(7) & 0.243(6) \\ 
\hline
 768 &  0.1309(4) & 0$^\ast$  & 0.76 & 1.784(9) & 0$^\ast$      & 0.01 &  2.05(1) & 0.216(9) \\ 
     &  0.1308(2) & -0.001(1) & 1.54 & 1.875(6) & $-$0.65(3)      & 0.00 & 2.137(7) & 0.125(6) \\ 
     &            &           &      & 1.767(9) & 0.125$^\ast$  & 0.01 &  2.03(1) & 0.233(9) \\ 
\hline
1024 &  0.1311(5) & 0$^\ast$  & 1.36 &  1.79(1) & 0$^\ast$      & 0.00 &  2.05(2) &  0.21(1) \\ 
     &            &           &      &  1.77(1) & 0.125$^\ast$  & 0.01 &  2.03(2) &  0.23(1) \\ 
\hline
\end{tabular}
\end{center}
\label{indices_Z7_Nt2}
\end{table}

\begin{table}
\caption{The same as Table~\ref{indices_Z5_Nt2} for $Z(7)$ with $N_t=4$,
determined at $\beta_c^{(1)} = 3.406$.}
\begin{center}
\setlength{\tabcolsep}{2.5pt}
\begin{tabular}{|c|c|c|c|c|c|c|c|c|}
\hline
$L_{\min}$ & $\beta/\nu$  & $r_{\beta/\nu}$  & $\chi^2_{\beta/\nu}$ 
           & $\gamma/\nu$ & $r_{\gamma/\nu}$ & $\chi^2_{\gamma/\nu}$ 
           & $d = 2 \beta/\nu + \gamma/\nu$  & $\eta = 2-\gamma/\nu$\\
\hline
 128 & 0.1294(1) & 0$^\ast$  & 4.64 & 1.772(3) & 0$^\ast$     & 1.72 & 2.031(4) & 0.228(3) \\ 
     &  0.141(2) &   0.07(1) & 0.54 & 1.766(4) &  0.04(2)     & 2.16 & 2.049(9) & 0.234(4) \\ 
     &           &           &      & 1.751(3) & 0.125$^\ast$ & 1.76 & 2.010(4) & 0.249(3) \\ 
\hline
 192 & 0.1298(2) & 0$^\ast$  & 3.61 & 1.771(5) & 0$^\ast$     & 2.12 & 2.031(5) & 0.229(5) \\ 
     & 0.1456(2) & 0.0961(8) & 0.20 &  1.85(8) & $-$0.5(5)    & 2.64 &  2.14(8) &  0.15(8) \\ 
     &           &           &      & 1.751(5) & 0.125$^\ast$ & 2.20 & 2.010(5) & 0.249(5) \\ 
\hline
 256 & 0.1302(2) & 0$^\ast$  & 1.43 & 1.778(6) & 0$^\ast$     & 1.84 & 2.038(6) & 0.222(6) \\ 
     & 0.1432(2) & 0.0811(9) & 0.19 &  1.74(3) &   0.2(2)     & 2.58 &  2.03(3) &  0.26(3) \\ 
     &           &           &      & 1.758(6) & 0.125$^\ast$ & 1.77 & 2.018(6) & 0.242(6) \\ 
\hline
 384 & 0.1307(3) & 0$^\ast$  & 0.37 & 1.773(9) & 0$^\ast$     & 2.48 &  2.03(1) & 0.227(9) \\ 
     & 0.1303(2) & $-$0.003(1) & 0.78 & 1.871(7) & $-$0.63(3) & 5.32 & 2.132(7) & 0.129(7) \\ 
     &           &           &      & 1.754(9) & 0.125$^\ast$ & 2.45 &  2.01(1) & 0.246(9) \\ 
\hline
 512 & 0.1310(5) & 0$^\ast$  & 0.08 &  1.78(1) & 0$^\ast$     & 4.86 &  2.04(1) &  0.22(1) \\ 
     &           &           &      &  1.76(1) & 0.125$^\ast$ & 4.77 &  2.02(1) &  0.24(1) \\ 
\hline
\end{tabular}
\end{center}
\label{indices_Z7_Nt4}
\end{table}

The procedure for the determination of the critical indices at the second
transition is similar to the one for the first transition, with the difference
that the fit with the scaling laws Eqs.~(\ref{magn_fss}) and~(\ref{chiM_fss}) 
is to be applied to data of the rotated magnetization, $M_R$, and of 
its susceptibility, $\chi_L^{(M_R)}$, respectively.
As the value of the critical coupling $\beta^{(2)}_{\rm c}$ we use the one
determined from the method (e) (see Table~\ref{crit_betas}). 

The results of the fits are summarized in 
Tables~\ref{indices_Z5_Nt2_2},~\ref{indices_Z5_Nt4_2},~\ref{indices_Z7_Nt2_2},~\ref{indices_Z7_Nt4_2}, for the cases of $Z(5)$ with $N_t$=2, 
$Z(5)$ with $N_t$=4, $Z(7)$ with $N_t$=2 and $Z(7)$ with $N_t$=4, 
respectively. The reference value for the index $\eta$ at this transition is 
$4/N^2$, {\it i.e.} $\eta=0.16$ for $N=5$ and $\eta=0.0816..$ for $N=7$,
whereas the hyperscaling relation to be fulfilled is $\gamma/\nu+2\beta/\nu=d
=2$.

A general comment is that in many of the cases we investigated both $d$ and 
$\eta$ at the two critical points slightly differ from the expected values,
though these differences cancel to a large extent if we define $\eta$ as 
$2 \beta/\nu$.

\begin{table}
\caption{Critical indices $\beta/\nu$ and $\gamma/\nu$ for the second 
transition in $Z(5)$ with $N_t$ = 2, determined by the fits given in 
Eqs.~(\ref{magn_fss}) and~(\ref{chiM_fss}) on the rotated magnetization 
$M_R$ and its susceptibility $\chi_L^{(M_R)}$ at $\beta_{\rm c}^{(2)}= 1.940$ 
for different choices of the minimum lattice size $L_{\rm min}$
(an asterisk indicates a fixed parameter).}
\begin{center}
\setlength{\tabcolsep}{2.5pt}
\begin{tabular}{|c|c|c|c|c|c|c|c|c|}
\hline
$L_{\min}$ & $\beta/\nu$  & $r_{\beta/\nu}$  & $\chi^2_{\beta/\nu}$ 
           & $\gamma/\nu$ & $r_{\gamma/\nu}$ & $\chi^2_{\gamma/\nu}$ 
           & $d = 2 \beta/\nu + \gamma/\nu$ & $\eta = 2 - \gamma/\nu$\\
\hline
 128 & 0.037(8)   & 0$^\ast$  & 2.22 & 1.918(2) & 0$^\ast$ & 0.52 & 1.99(2) & 0.082(2) \\ 
     & $-$0.27(3) & $-$1.8(2) & 1.07 & 1.900(3) & 0.11(1) & 0.53 & 1.36(7) & 0.100(3) \\ 
\hline
 192 &  0.03(1)   & 0$^\ast$  & 2.10 & 1.918(3) & 0$^\ast$ & 0.64 & 1.97(2) & 0.082(3) \\ 
     & $-$0.25(4) & $-$1.6(2) & 1.52 & 1.821(3) & 0.59(1) & 0.31 & 1.33(8) & 0.179(3) \\ 
\hline
 256 &  0.01(1)   & 0$^\ast$  & 2.27 & 1.916(4) & 0$^\ast$ & 0.64 & 1.94(3) & 0.084(4) \\ 
     & $-$0.20(5) & $-$1.3(3) & 2.38 & 1.900(3) & 0.10(1) & 0.82 &  1.5(1) & 0.100(3) \\ 
\hline
 384 & $-$0.02(2) & 0$^\ast$  & 1.77 & 1.913(6) & 0$^\ast$ & 0.62 & 1.88(5) & 0.087(6) \\ 
     & $-$0.01(2) & 0.03(6)   & 3.56 &  1.87(3) &  0.3(2) & 1.06 & 1.85(6) &  0.13(3) \\ 
\hline
 512 & $-$0.05(3) & 0$^\ast$  & 1.37 & 1.912(8) & 0$^\ast$ & 1.21 & 1.81(7) & 0.088(8) \\ 
\hline
\end{tabular}
\end{center}
\label{indices_Z5_Nt2_2}
\end{table}

\begin{table}
\caption{The same as Table~\ref{indices_Z5_Nt2_2} for $Z(5)$ with $N_t=4$,
determined at $\beta_c^{(2)} =2.898$.}
\begin{center}
\setlength{\tabcolsep}{2.5pt}
\begin{tabular}{|c|c|c|c|c|c|c|c|c|}
\hline
$L_{\min}$ & $\beta/\nu$  & $r_{\beta/\nu}$  & $\chi^2_{\beta/\nu}$ 
           & $\gamma/\nu$ & $r_{\gamma/\nu}$ & $\chi^2_{\gamma/\nu}$ 
           & $d = 2 \beta/\nu + \gamma/\nu$ & $\eta = 2 - \gamma/\nu$\\
\hline
 128 &  0.17(1) & 0$^\ast$ & 1.52 & 1.850(3) & 0$^\ast$ & 0.07 & 2.20(3) & 0.150(3) \\ 
     & $-$0.28(5) &  $-$2.6(3) & 0.30 & 1.847(4) & 0.02(2)  & 0.08 &  1.3(1) & 0.153(4) \\ 
\hline
 192 &  0.15(1) & 0$^\ast$ & 0.40 & 1.849(4) & 0$^\ast$ & 0.07 & 2.15(3) & 0.151(4) \\ 
     &  $-$0.1(1) &  $-$1.5(6) & 0.25 & 1.847(4) & 0.01(2)  & 0.09 &  1.6(2) & 0.153(4) \\ 
\hline
 256 &  0.15(2) & 0$^\ast$ & 0.51 & 1.849(5) & 0$^\ast$ & 0.07 & 2.14(4) & 0.151(5) \\ 
     & $-$0.09(9) &  $-$1.5(5) & 0.36 & 1.846(5) & 0.01(2)  & 0.10 &  1.7(2) & 0.154(5) \\ 
\hline
 384 &  0.12(3) & 0$^\ast$ & 0.24 & 1.847(8) & 0$^\ast$ & 0.07 & 2.09(7) & 0.153(8) \\ 
     &  0.10(2) & $-$0.15(9) & 0.45 & 1.845(6) & 0.01(2) & 0.14 & 2.04(5) & 0.155(6) \\ 
\hline
 512 &  0.10(5) & 0$^\ast$ & 0.21 &  1.85(1) & 0$^\ast$ & 0.11 &  2.1(1) &  0.15(1) \\ 
\hline
\end{tabular}
\end{center}
\label{indices_Z5_Nt4_2}
\end{table}

\begin{table}
\caption{The same as Table~\ref{indices_Z5_Nt2_2} for $Z(7)$ with $N_t=2$,
determined at $\beta_c^{(2)} = 3.366$.}
\begin{center}
\setlength{\tabcolsep}{2.5pt}
\begin{tabular}{|c|c|c|c|c|c|c|c|c|}
\hline
$L_{\min}$ & $\beta/\nu$  & $r_{\beta/\nu}$  & $\chi^2_{\beta/\nu}$ 
           & $\gamma/\nu$ & $r_{\gamma/\nu}$ & $\chi^2_{\gamma/\nu}$ 
           & $d = 2 \beta/\nu + \gamma/\nu$ & $\eta = 2 - \gamma/\nu$\\
\hline
 128 & 0.034(6) & 0$^\ast$ & 2.33 & 1.921(2) & 0$^\ast$ & 0.45 & 1.99(1) & 0.079(2) \\ 
     & $-$0.27(5) & $-$1.9(3)  & 0.34 &  1.89(3) &  0.2(2)  & 0.27 &  1.3(1) &  0.11(3) \\ 
\hline
 192 & 0.018(7) & 0$^\ast$ & 0.72 & 1.919(2) & 0$^\ast$ & 0.23 & 1.96(2) & 0.081(2) \\ 
     & $-$0.19(4) & $-$1.3(3)  & 0.26 & 1.903(2) & 0.10(1)  & 0.26 & 1.53(9) & 0.097(2) \\ 
\hline
 256 & 0.010(9) & 0$^\ast$ & 0.26 & 1.919(2) & 0$^\ast$ & 0.28 & 1.94(2) & 0.081(2) \\ 
     & $-$0.12(7) & $-$0.8(4)  & 0.18 & 1.904(3) & 0.10(1)  & 0.28 &  1.7(1) & 0.096(3) \\ 
\hline
 384 &     0(1) & 0$^\ast$ & 11.54& 1.919(3) & 0$^\ast$ & 0.35 &    2(2) & 0.081(3) \\ 
     & $-$0.12(6) & $-$0.9(4)  & 0.21 & 1.904(3) & 0.10(1)  & 0.37 &  1.7(1) & 0.096(3) \\ 
\hline
 512 &  0.00(2) & 0$^\ast$ & 0.25 & 1.917(4) & 0$^\ast$ & 0.21 & 1.92(3) & 0.083(4) \\ 
     &  0.00(1) & 0.00(5)  & 0.38 & 1.903(3) & 0.10(1)  & 0.29 & 1.91(3) & 0.097(3) \\ 
\hline
 768 & $-$0.01(2) & 0$^\ast$ & 0.11 & 1.916(6) & 0$^\ast$ & 0.29 & 1.90(5) & 0.084(6) \\ 
     & $-$0.01(1) & 0.02(6)  & 0.22 & 1.903(4) & 0.10(2)  & 0.56 & 1.89(3) & 0.097(4) \\ 
\hline
1024 & $-$0.00(3) & 0$^\ast$ & 0.14 & 1.918(8) & 0$^\ast$ & 0.53 & 1.91(7) & 0.082(8) \\ 
\hline
\end{tabular}
\end{center}
\label{indices_Z7_Nt2_2}
\end{table}

\begin{table}
\caption{The same as Table~\ref{indices_Z5_Nt2_2} for $Z(7)$ with $N_t=4$,
determined at $\beta_c^{(2)} = 5.158$.}
\begin{center}
\setlength{\tabcolsep}{2.5pt}
\begin{tabular}{|c|c|c|c|c|c|c|c|c|}
\hline
$L_{\min}$ & $\beta/\nu$  & $r_{\beta/\nu}$  & $\chi^2_{\beta/\nu}$ 
           & $\gamma/\nu$ & $r_{\gamma/\nu}$ & $\chi^2_{\gamma/\nu}$ 
           & $d = 2 \beta/\nu + \gamma/\nu$  & $\eta = 2 - \gamma/\nu$\\
\hline
 128 & 0.037(8) & 0$^\ast$ & 2.22 & 1.918(2) & 0$^\ast$ & 0.52 & 1.99(2) & 0.082(2) \\ 
     & $-$0.27(3) & $-$1.8(2)  & 1.07 & 1.900(3) & 0.11(1)  & 0.53 & 1.36(7) & 0.100(3) \\ 
\hline
 192 &  0.03(1) & 0$^\ast$ & 2.10 & 1.918(3) & 0$^\ast$ & 0.64 & 1.97(2) & 0.082(3) \\ 
     & $-$0.25(4) & $-$1.6(2)  & 1.52 & 1.821(3) & 0.59(1)  & 0.31 & 1.33(8) & 0.179(3) \\ 
\hline
 256 &  0.01(1) & 0$^\ast$ & 2.27 & 1.916(4) & 0$^\ast$ & 0.64 & 1.94(3) & 0.084(4) \\ 
     & $-$0.20(5) & $-$1.3(3)  & 2.38 & 1.900(3) & 0.10(1)  & 0.82 &  1.5(1) & 0.100(3) \\ 
\hline
 384 & $-$0.02(2) & 0$^\ast$ & 1.77 & 1.913(6) & 0$^\ast$ & 0.62 & 1.88(5) & 0.087(6) \\ 
     & $-$0.01(2) & 0.03(6)  & 3.56 &  1.87(3) &  0.3(2)  & 1.06 & 1.85(6) &  0.13(3) \\ 
\hline
 512 & $-$0.05(3) & 0$^\ast$ & 1.37 & 1.912(8) & 0$^\ast$ & 1.21 & 1.81(7) & 0.088(8) \\ 
\hline
\end{tabular}
\end{center}
\label{indices_Z7_Nt4_2}
\end{table}

\begin{figure}[t]
\includegraphics[width=0.45\textwidth]{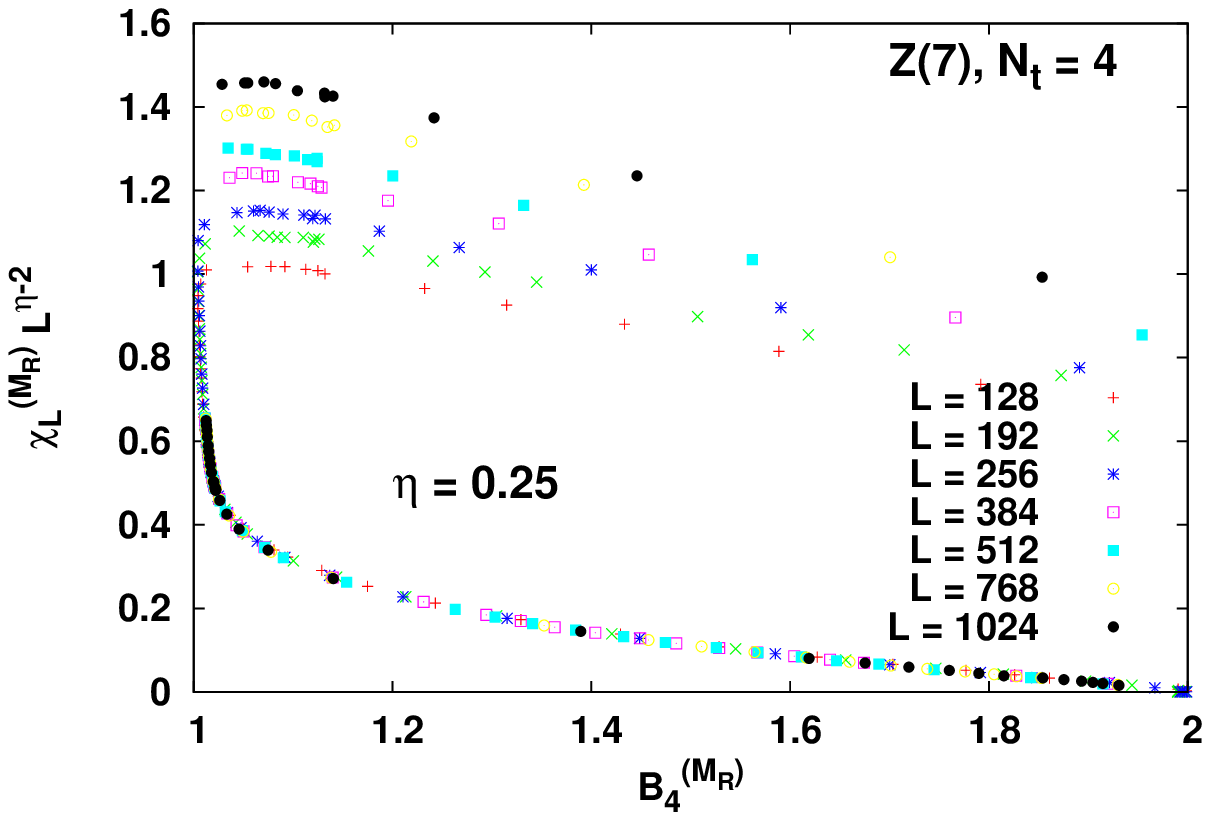}
\includegraphics[width=0.45\textwidth]{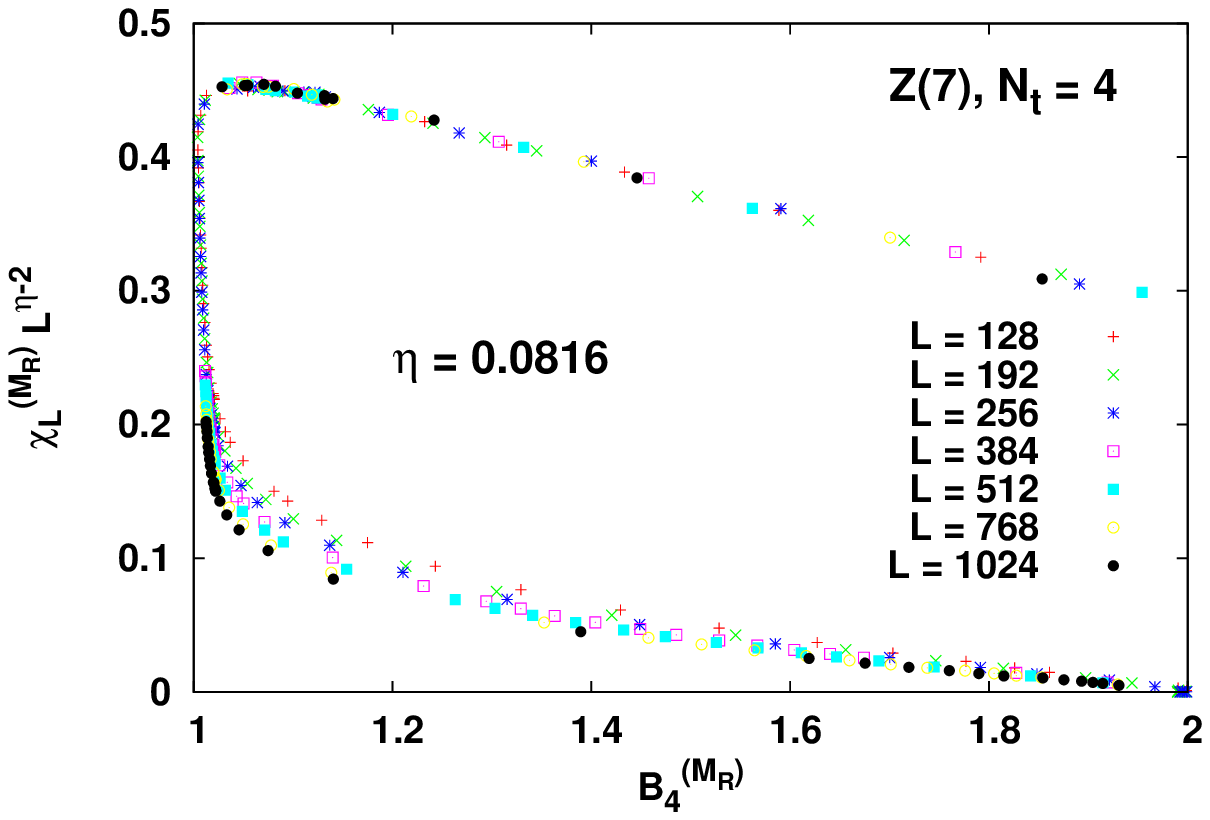}
\caption{Correlation between $\chi_{L}^{(M_{R})}L^{\eta-2}$ and 
the Binder cumulant $B_{4}^{(M_{R})}$ in $Z(7)$ with $N_t=4$ for $\eta=0.25$ 
(left) and for $\eta=0.0816$ (right) on lattices with $L$ ranging from 128
to 1024.}
\label{CHIvsB4z7_2_nt4}
\end{figure}

\begin{figure}
\includegraphics[width=0.45\textwidth]{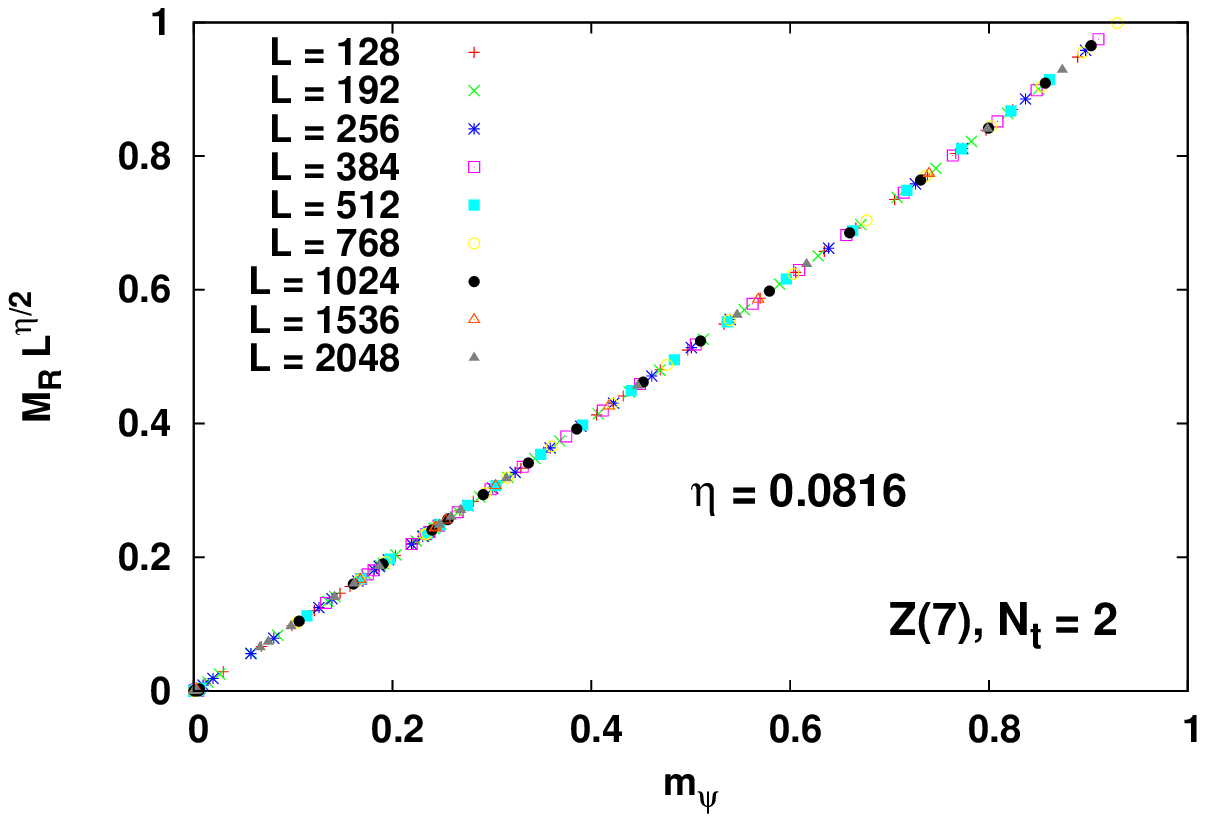}
\includegraphics[width=0.45\textwidth]{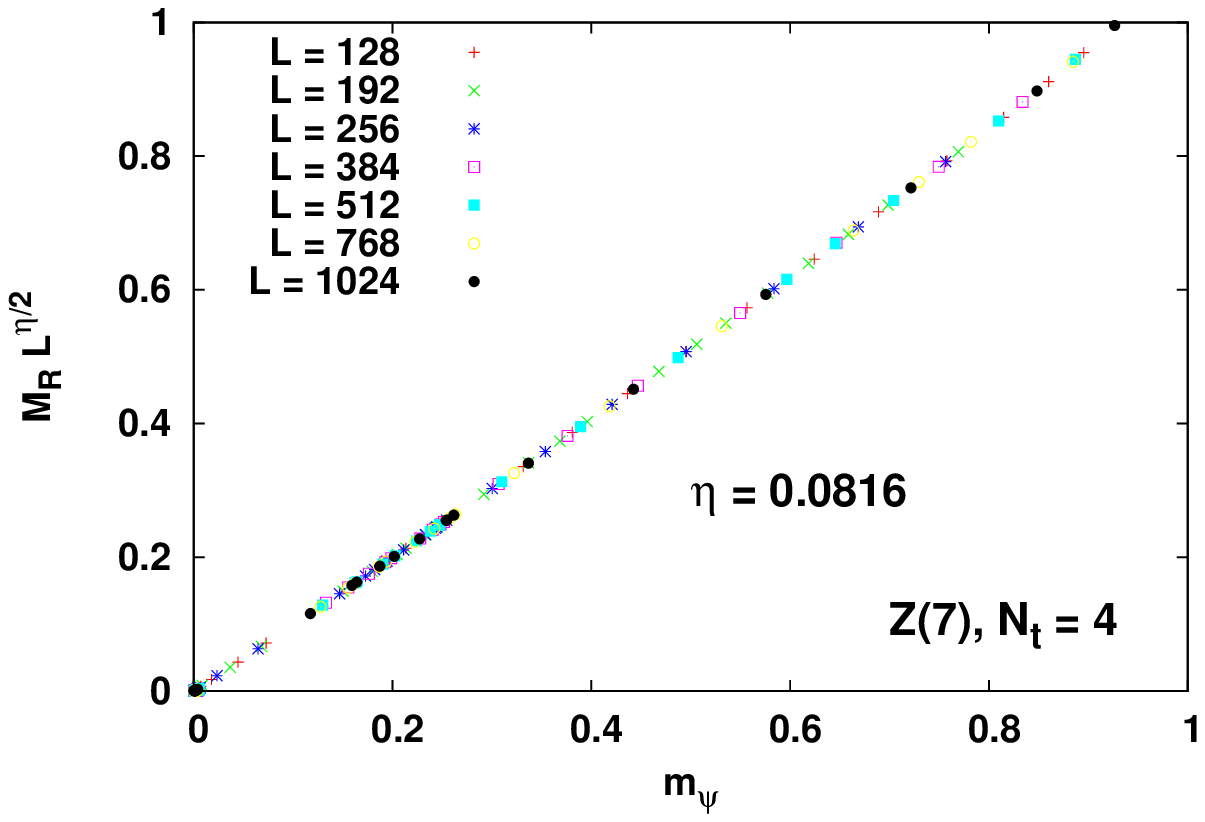}
\caption{Correlation between $M_{R}L^{\eta/2}$ and $m_\psi$
in $Z(7)$ with $N_t=2$ (left) and in $Z(7)$ with $N_t=4$ (right)
for $\eta=0.0816$ on lattices with various values of $L$.}
\label{MRLvMPSIz7_2}
\end{figure}

There is an independent method to determine the critical exponent $\eta$,
which does not rely on the prior knowledge of the critical coupling, but 
is based on the construction of a suitable universal 
quantity~\cite{Loison99,2dzn}. The idea is to plot 
$\chi_{L}^{(M_{R})}L^{\eta-2}$ versus $B_{4}^{(M_{R})}$ and to look for 
the value of $\eta$ which optimizes the overlap of curves from different 
volumes. This method is illustrated in Fig.~\ref{CHIvsB4z7_2_nt4} for the case
of $Z(7)$ with $N_t=4$: for $\eta=0.25$ the overlap is optimal in the lower
branch of the curves, corresponding to the region of the first transition,
while per $\eta=0.0816\simeq 4/7^2$ the overlap is optimal in the upper
branch. Another option is to plot $M_{R}L^{\eta/2}$ versus $m_\psi$,
which leads to overlapping curves for $\eta$ fixed at the value of the
second phase transition, as illustrated in Fig.~\ref{MRLvMPSIz7_2}
for the case of $Z(7)$ with $N_t=2$ and $N_t=4$. 

\begin{figure}
\includegraphics[width=0.45\textwidth]{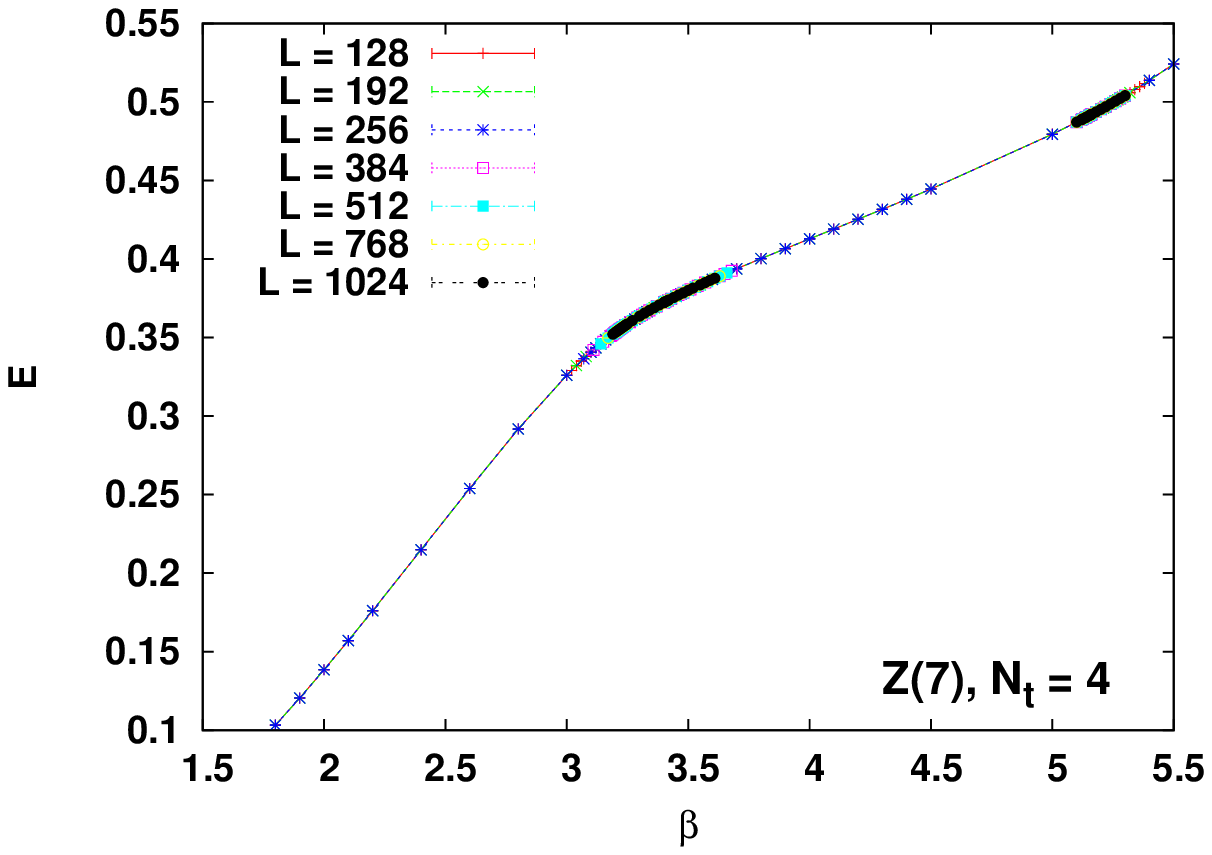}
\includegraphics[width=0.45\textwidth]{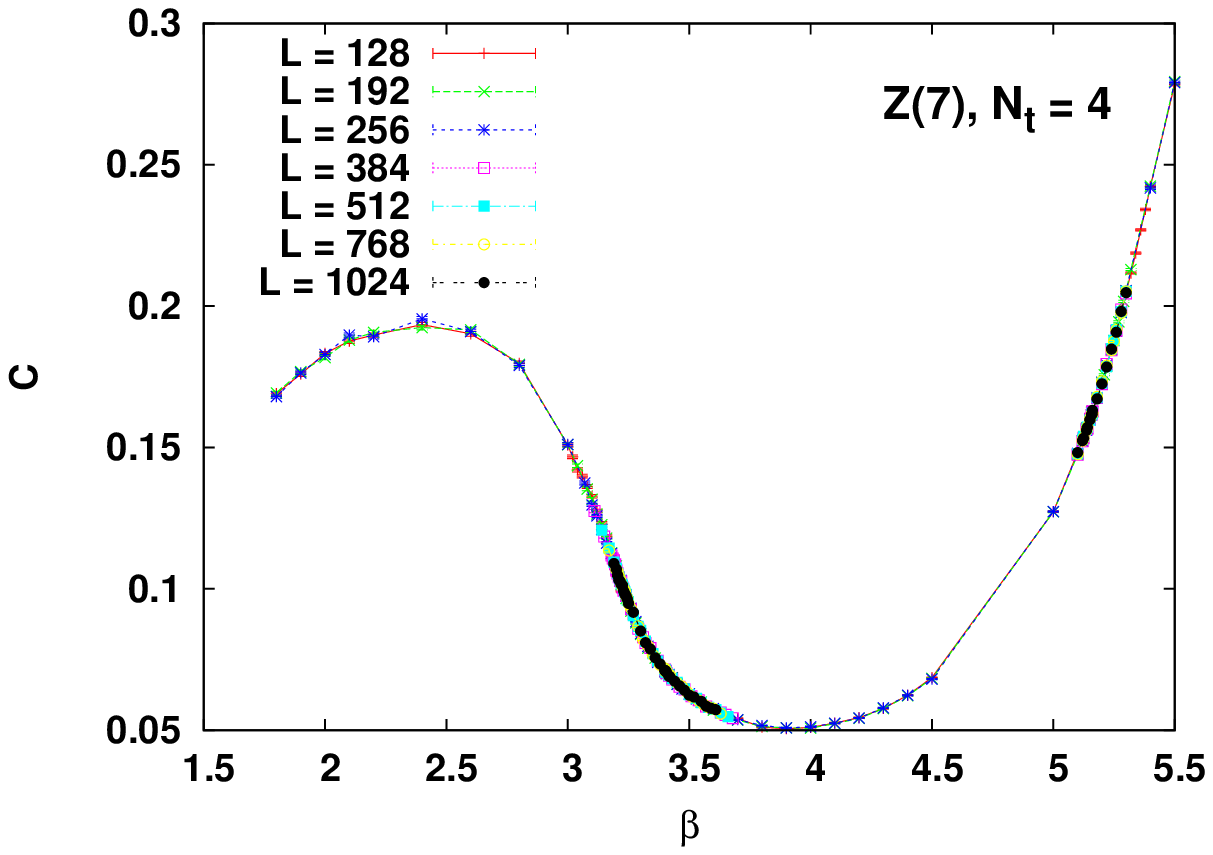}
\caption{Mean link energy (left) and specific heat (right) versus 
$\beta$ in $Z(7)$ with $N_t = 4$ on lattices with $L$ ranging from 
128 to 1024.}
\label{fig:spheat_Z7_NT4}
\end{figure}

\clearpage

\subsection{Other checks of the nature of the phase transitions}

To produce further evidence in favor of the fact that the phase transitions 
investigated so far are both of infinite order, we have calculated
the mean link energy $E$ and the specific heat $C$ around the transitions
in $Z(5)$ and in $Z(7)$ with $N_t$=2 and $N_t=4$ (see 
Fig.~\ref{fig:spheat_Z7_NT4} for the case of $Z(7)$ with $N_t=4$, for 
example). In all cases the dependence of $E$ and $C$ on $\beta$ is continuous, 
so that first and second order transitions are ruled out.
 
\section{Behavior with $N$ of the critical couplings}

The results of this work and those available in the literature allow us to
make some considerations about the behavior with $N$ of the critical couplings 
$\beta_{\rm c}^{(1)}$ and $\beta_{\rm c}^{(2)}$. Examining our data for 
$\beta_{\rm c}^{(1,2)}$, one concludes that, for a fixed $N_t$,
\begin{itemize}
\item 
$\beta_{\rm c}^{(1)}$ converges to the $XY$ value very fast, like 
$\exp(-a N^2)$ \ ,
\item 
$\beta_{\rm c}^{(2)}$ diverges like $N^2$ \;.
\end{itemize}

To better see the dependence on $N$ of the critical couplings,  
we have found also the critical coupling $\beta_{\rm c}^{(1,2)}$ in
$Z(9)$ and $Z(13)$ with $N_t=2$ and $N_t=4$, using the method (b)
for $\beta_{\rm c}^{(1)}$ and the method (e) for $\beta_{\rm c}^{(2)}$ 
(see Section~\ref{sec:betas} for the details of these methods).
In Table~\ref{Ndep} we summarize the present knowledge about the position of 
the critical points for $3d$ $Z(N)$ gauge models at $\beta_s=0$. 

\begin{table}[tb]\setlength{\tabcolsep}{12pt}
\centering
\caption[]{Summary of the known values of the critical couplings 
$\beta_{\rm c}^{(1)}$ and $\beta_{\rm c}^{(2)}$ in $Z(N)$ with $N_t=2,4$.}
\vspace{0.2cm}
\begin{tabular}{|c|c|l|l|c|}
\hline
$N$ & $N_t$ & \hspace{0.5cm}$\beta_{\rm c}^{(1)}$ 
&\hspace{0.7cm}$\beta_{\rm c}^{(2)}$ & Reference \\
\hline
 5       & 2 & 1.87(1)  & \hspace{0.11cm} 1.940(7)  & this work  \\
 7       & 2 & 2.031(7) & \hspace{0.11cm} 3.366(7)  & this work  \\
 9       & 2 & 2.04(3)  & \hspace{0.11cm} 5.38(4)   & this work  \\
13       & 2 & 2.02(1)  & \hspace{0.11cm} 10.815(8) & this work  \\
$\infty$ & 2 & -- & \hspace{0.7cm}  $\infty$  &            \\
\hline
 5       & 4 & 2.813(3) & \hspace{0.11cm} 2.898(4)  & this work     \\
 7       & 4 & 3.406(8) & \hspace{0.11cm} 5.158(7)  & this work     \\
 9       & 4 & 3.50(1)  & \hspace{0.11cm} 8.28(1)   & this work     \\
13       & 4 & 3.490(6) & \hspace{0.11cm} 16.94(2)  & this work     \\
$\infty$ & 4 & 3.42(1)  & \hspace{0.7cm} $\infty$   & \cite{3du1ft} \\
\hline
\end{tabular}
\label{Ndep}
\end{table}

One should expect that the $3d$ $Z(N)$ gauge models at $\beta_s=0$ satisfy 
the scaling predicted by RG, probably up to ${\cal{O}}(N)$ corrections. 
One could try therefore to fit the available Monte Carlo data for 
$\beta_{\rm c}^{(1,2)}$ with formulae predicted by RG and modified to account 
for such corrections.  

We find that the critical couplings for the first transition are well
reproduced by the function
\[
\beta_{\rm c}^{(1)} = A  - B N^2 \exp {\left( - \frac{N^2}{C} \right)}\;,
\]
for suitable values of the parameters $A$, $B$ and $C$, both for $N_t=2$
and $N_t=4$.

The critical couplings for the second transition are well reproduced instead
by the following functions:
\[
\beta_{\rm c}^{(2)} = A N^2 + B N + C
\]
and
\[
\beta_{\rm c}^{(2)} = \frac{A}{1-\cos\left(\frac{2 \pi}{N}\right)} + B N + C\;,
\]
for suitable values of the parameters $A$, $B$ and $C$, both for $N_t=2$
and $N_t=4$. The last formula has been suggested in Ref.~\cite{bhanot} in the context 
of the zero-temperature theory. 

\begin{figure}
\includegraphics[width=0.45\textwidth]{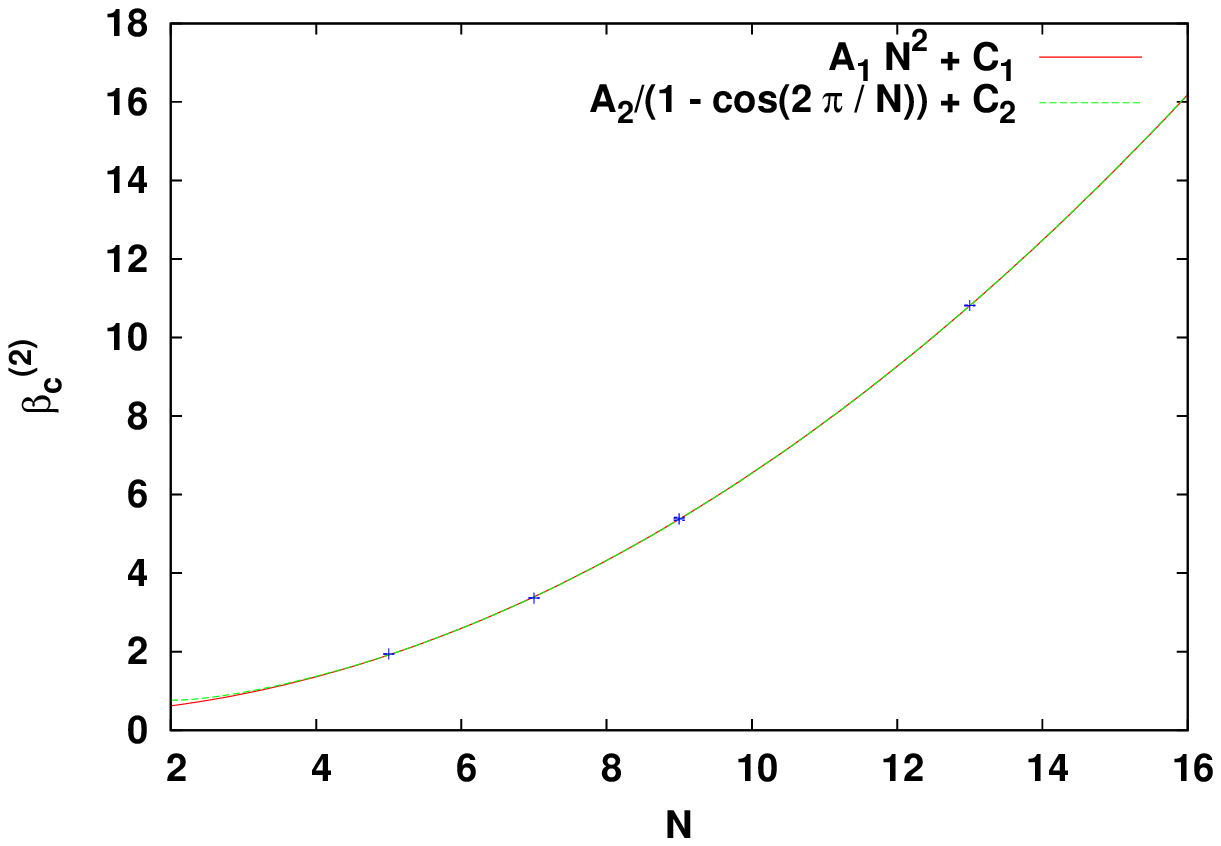}
\includegraphics[width=0.45\textwidth]{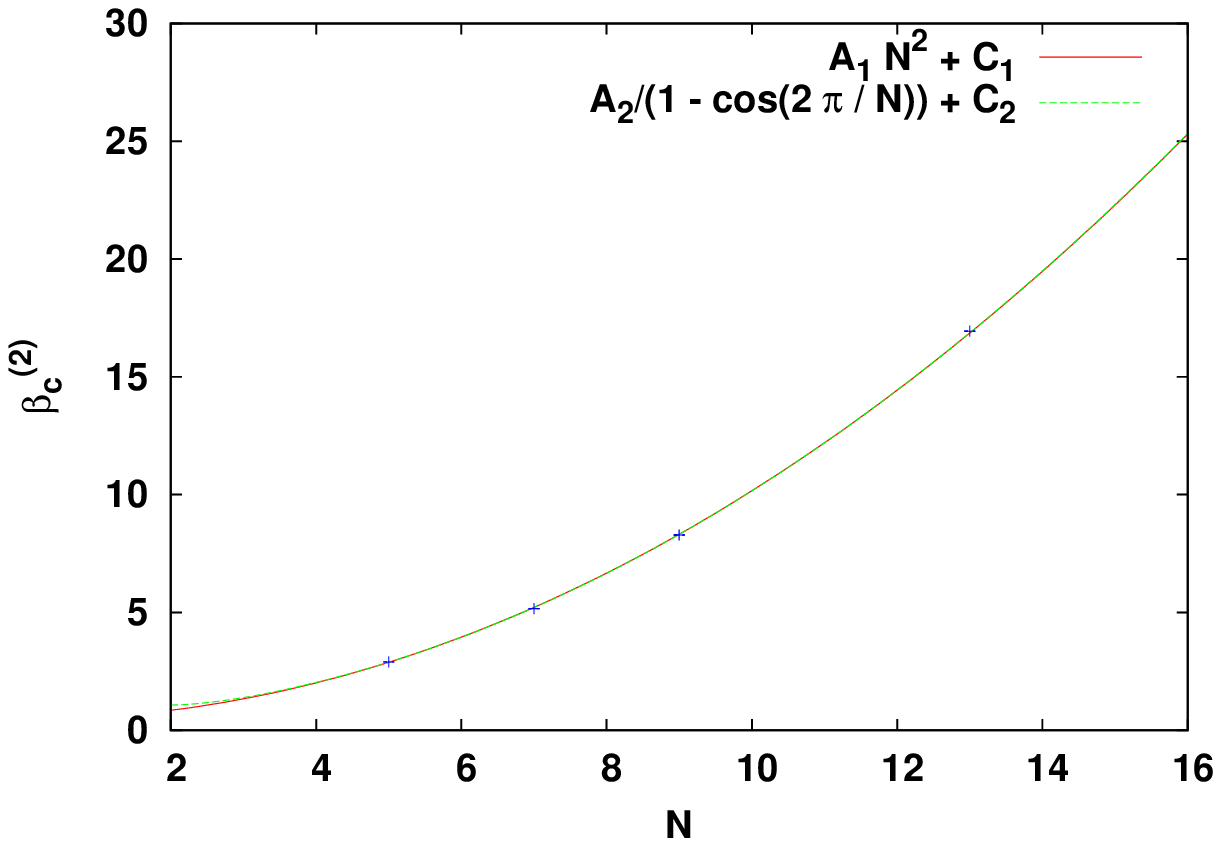}
\caption{Fits of critical values for $\beta_{\rm c}^{(2)}$ as a function of $N$ 
for $N_t=2$ (left) and $N_t=4$ (right).}
\label{fig:scaling2}
\end{figure}

As an illustration of our fits, Fig.~\ref{fig:scaling2} shows the dependence of 
$\beta_{\rm c}^{(2)}$ on $N$ (parameter $B=0$ in fitting formulas).
One concludes from these plots that to distinguish 
between two scalings one should probably have data for smaller values of $N=2,3,4$ 
at $\beta_s=0$.

\section{Conclusions} 

In this paper we have studied the $3d$ $Z(N)$ gauge theory at the finite 
temperature in the strong coupling region $\beta_s=0$. This study was based 
on the exact relation of this model to a generalized $2d$ $Z(N)$ spin model. 
In Section~2 we established the exact relation between couplings of these two 
models, described qualitatively some of the RG predictions for effective 
model and gave analytical estimations of the critical couplings. 
  
The main, numerical part of the work has been devoted to the localization
of the critical couplings and to the computation of the critical indices:
\begin{itemize}

\item We have determined numerically the two critical couplings of 
$Z(N=5,7,9,13)$ LGTs and given estimates of the critical indices $\eta$
at both transitions. For the first time we have a clear indication that for 
all $N\geq 5$ the scenario of three phases is realized: 
a disordered phase at small $\beta_t$, a massless or BKT one at intermediate 
values of $\beta_t$ and an ordered phase, occurring at larger and larger 
values of $\beta_t$ as $N$ increases.  
This matches perfectly with the $N\to\infty$ limit, {\it i.e.} the 
finite-temperature $3d$ $U(1)$ LGT (at $\beta_s=0$), where the ordered phase 
is absent;

\item We have found that the values of the critical index $\eta$ at the two 
transitions are compatible with the theoretical expectations; 
in order to reproduce the expected value of $\eta$, we had to take into account 
logarithmic corrections with the exponent $r$ fixed to 0.125;

\item The index $\nu$ also appears to be compatible with the value $1/2$, 
in agreement with RG predictions. 
\end{itemize}

Results listed above present further evidence that finite-temperature
$3d$ $Z(N)$ LGTs for $N>4$ undergo two phase transitions of the BKT type.
 
Moreover, this model belongs to the universality class of the $2d$ $Z(N)$ 
spin model, at least in the strong coupling limit $\beta_s=0$.

Considering the determinations of the critical couplings as a function of $N$, 
we have conjectured the approximate scaling for $\beta_c^{(1,2)}(N)$. 

\begin{figure}
\centering
\includegraphics[width=0.5\textwidth]{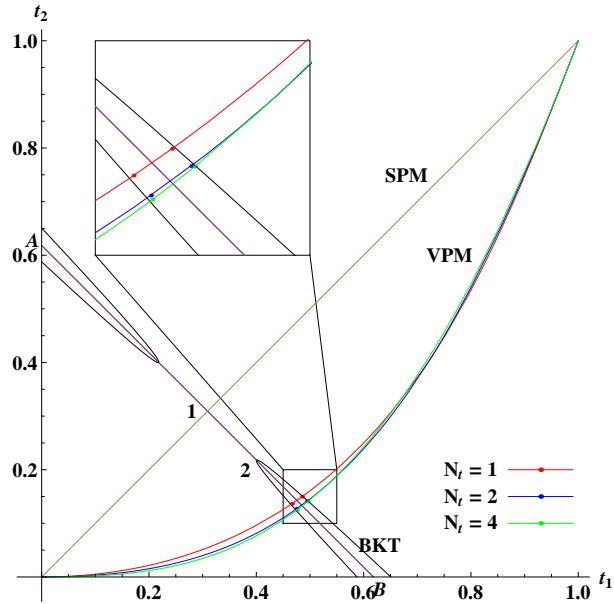}
\caption{Phase structure of the general $Z(5)$ spin model (see text 
for explanation).}
\label{fig:phasez5}
\end{figure}

Finally, the study performed here allows to improve our knowledge of the 
phase diagrams of the generalized $2d$ $Z(N)$ spin models. As an example, we 
plot in Fig.~\ref{fig:phasez5} the general phase diagram for $N=5$ in the 
$(t_1, t_2)$-plane, where $t_i = (B_i/B_0)^{N_t}$, and $B_i$ are 
defined in~(\ref{couplings_coeff}). Here, the line $AB$ is self-dual line, 
SPM corresponds to the standard Potts model, VPM to the vector model.
The SPM undergoes a first order phase transition with a critical point 
occurring on the self-dual line. The line of the first order phase transition 
terminates at the point 2. Its approximate position was computed in 
Ref.~\cite{Domany}. Shown are also the locations of the critical points 
for the VPM ($N_t=1$) and for $Z(5)$ LGT with $N_t=2,4$. 
The parametric curves for different $N_t$ lie very close to each other, 
so we cannot trace a sufficiently big part of the curve while changing $N_t$. 
On the other hand it shows that already the model with $N_t=4$ presents a very good 
approximation to the finite-temperature limit. Indeed, the parametric curve for 
$N_t=8$ is almost indistinguishable from the curve with $N_t=4$.

\section{Acknowledgments}

The work of O.B. was supported by the Program of Fundamental
Research of the Department of Physics and Astronomy of NAS, Ukraine.
The work of G.C. and M.G. was supported in part by the European Union
un\-der ITN STRO\-NG\-net (grant PITN-GA-2009-238353). G.C. thanks Matteo
Giordano for useful discussions.

\end{document}